\documentclass[a4paper,10pt]{article}
\usepackage[position=t,singlelinecheck=off]{subfig}
\usepackage{amssymb,amsmath,graphics,graphicx,float,braket}
\usepackage{array,multirow,setspace,makecell,pbox}
\usepackage[colorlinks=true, citecolor=blue, urlcolor=blue ]{hyperref}
\usepackage{cite}
\usepackage[title]{appendix}
\def\beq{\begin{equation}}
\def\eeq{\end{equation}}
\def\bea{\begin{eqnarray}}
\def\eea{\end{eqnarray}}
\def\nn{\nonumber}

\makeatletter
\def\@cite#1#2{${\mbox{#1\if@tempswa , #2\fi}}$}
\makeatother

\renewcommand{\thesection}{\arabic{section}}

\newcommand\norm[1]{\left\lVert#1\right\rVert}
\newcolumntype{P}[1]{>{\centering\arraybackslash}p{#1}}
\newcolumntype{M}[1]{>{\centering\arraybackslash}m{#1}}
\begin{document}
\thispagestyle{empty}
\begin{center}
\begin{LARGE}
\textsf{Quantum coherence, correlations and nonclassical states in the two-qubit Rabi model with parametric oscillator}
\end{LARGE} \\

\bigskip\bigskip
V. Yogesh$^{\dagger}$ and Prosenjit Maity$^{*}$ 
\\
\begin{small}
\bigskip
\textit{
 $^{\dagger}$ Department of Theoretical Sciences, S. N. Bose National Centre for Basic Sciences, \\ Block-JD, Sector-III, Salt Lake, Kolkata 700106, India. \\
$^{*}$ Department of Physics, Ramakrishna Mission Residential College, \\ Narendrapur, Kolkata-700103, India.}
\end{small}
\end{center}

\vfill
\begin{abstract}
Quantum coherence and quantum correlations are studied in the strongly interacting system composed of two qubits and an oscillator with the presence of a parametric medium. To analytically solve the system, we employ the adiabatic approximation approach. It assumes each qubit’s characteristic frequency is substantially lower than the oscillator frequency. To validate our approximation,  a good agreement between the calculated energy spectrum of the Hamiltonian with its numerical result is presented. The time evolution of the reduced density matrices of the two-qubit and the oscillator subsystems are computed from the tripartite initial state. Starting with a factorized two-qubit initial state, the quasi-periodicity in the revival and collapse phenomenon that occurs in the two-qubit population inversion is studied. Based on the measure of relative entropy of coherence, we investigate the quantum coherence and its explicit dependence on the parametric term both for the two-qubit and the individual qubit subsystems by adopting different choices of the initial states. Similarly, the existence of quantum correlations is demonstrated by studying the geometric discord and concurrence. Besides, by numerically minimizing the Hilbert-Schmidt distance,  the dynamically produced near maximally entangled states are reconstructed. The reconstructed states are observed to be nearly pure generalized Bell states. Furthermore, utilizing the oscillator density matrix,  the quadrature variance and phase-space distribution of the associated Husimi $Q$-function  are computed in the minimum entropy regime and conclude that the obtained nearly pure evolved state is a squeezed coherent state.

\end{abstract} 
 
\newpage
\setcounter{page}{1}

\section{Introduction}
Quantum correlations, as a fundamental property of a multipartite quantum system and an indispensable resource for quantum information processing [\cite{nielsen2002}], were initially investigated in the entanglement-versus-separability scenario [\cite{werner1989,horodecki2009,amico2008}]. Even though entanglement has received much attention from many authors, it is not a unique attribute of a quantum system that facilitates information tasks. There are cases [\cite{datta2008,lanyon2008}], even if there is no entanglement, still, quantum information processing tasks can  efficiently be performed by employing quantum discord [\cite{ollivier2001, henderson2001,zurek2003quantum}], which is supposed to be more feasible than entanglement. Quantum discord quantifies quantum correlations that exist beyond the entanglement i.e. there might be nonvanishing quantum discord even in the absence of entanglement [\cite{ollivier2001}]. Since the computation of quantum discord entails a complicated optimization method, generally it is difficult to derive the analytical results except for a few common examples of two-qubit cases [\cite{ali2010}]. It  has already been reported that the running time of any method for numerically computing quantum discord  is anticipated to rise exponentially with the Hilbert space dimensions. As a result, even with a reasonable scale, calculating quantum discord is difficult in practice [\cite{huang2014}].

\par
Given the complexity in estimating quantum discord, the  geometric measure of quantum discord (also known as geometric discord) has been introduced and an analytic formula for two-qubit systems was derived [\cite{vedral2010}]. Subsequently, an alternative approach  of  geometric discord was provided for a qubit-qudit system [\cite{tufarelli2012}]. In the light-matter interacting system,  geometric discord was studied in Jaynes-Cummings model consisting of atoms inside a cavity with an isolated atom [\cite{qiang2015}]. In addition to  quantum discord, quantum coherence [\cite{baumgratz2014}], which arises from the quantum superposition between different states of a quantum system, is one of the  fundamental resources in quantum information processing and quantum computation [\cite{horodecki2009,plenio2017,winter2016}]. 
Recent studies indicate that the coherence in a quantum state plays an important role  in the fields of  quantum thermodynamics [\cite{lostaglio2015,cwiklinski2015,misra2016}], quantum biology [\cite{lloyd2011,levi2014}] etc. Based on the framework laid out in  Ref. [\cite{baumgratz2014}], some measures for the quantum coherence have been put forward, for example, relative entropy of coherence [\cite{baumgratz2014,zhang2016}],  $l_{1}$-norm of coherence [\cite{baumgratz2014}], and trace-distance measure of coherence [\cite{rana2016}]. In particular,  using relative entropy of coherence, the quantum coherence was investigated in the nonresonant Jaynes–Cummings model, where the atom is initially prepared in an incoherent mixed state and the quantized field is in a thermocoherent state [\cite{rastegar2016}].

\par

 We consider two qubits interacting with a single-mode quantum field  in the  strong coupling domain in the Rabi model with the presence of a parametric oscillator. To explore the qubit-oscillator system under strong coupling strength where the oscillator frequency dominates the characteristic frequencies of the qubits, we employ  the adiabatic approximation approach [\cite{irish2005,ashhab2010}] that exploits the distinction between slow and rapidly varying degrees of freedom. It allows us to approximately diagonalize the entire Hamiltonian by decoupling its components corresponding to each time scale [\cite{irish2005}]. Using this approximation, physical systems consisting of two [\cite{yang2012,dong2016}] and three qubits [\cite{shen2014}] coupled with a single oscillator degree of freedom have already been studied. We construct the time evolution of the pure tripartite initial state, by considering the oscillator degree of freedom as a coherent state. By using the reduced density matrices of the qubits, the  quantum coherence for the two-qubit subsystem as well as its individual subsystems are studied. The quantum correlations are investigated by comparing the  geometric discord and concurrence for the initially factorized and entangled states and  the nonvanishing geometric discord is noticed at the entanglement sudden death  region [\cite{Ebarly2006}]. Furthermore, by initially starting with a factorized bipartite two-qubit subsystem,  the dynamically produced nearly pure generalized Bell states are obtained. On the other hand, by tracing over the qubit degrees of freedom, we derive the oscillator reduced density matrix and calculate the quadrature variance and Husimi $Q$-function to study the generated nearly pure squeezed coherent state at the minimum entropy configuration.
 
\par 
The work is organized as follows: In Sec.  \ref{sec_H}, the approximate diagonalization of the Hamiltonian is performed within the framework of adiabatic approximation. In Sec. \ref{dmatrix}, the time evolution of the reduced density matrices for the qubits and oscillator are obtained. In Sec. \ref{Rev_col}, we demonstrate  the revival and collapse phenomenon observed in the two-qubit population inversion. In Sec. \ref{cohen},  the quantum coherence of  the two-qubit subsystem and its constituent qubit subsystems  are investigated and the influence of parametric oscillator on them is illustrated. In Sec. \ref{Geo_dis}, for the measure of quantum correlations, the geometric discord and the concurrence are discussed and impact of parametric oscillator on them is shown. In Sec. \ref{nonclassical}, the generation of nonclassical states are studied. Sec. \ref{sec_con} contains the summary and conclusion of the work.

\section{Diagonalization of the Hamiltonian via adiabatic approximation}
\label{sec_H}
The two-qubit Rabi Hamiltonian [\cite{peng2012,BMLara,peng2014,duan2015,zhang2015,mao2015,mao2019,dong2016,zhang2021,yan2021}] in the presence of a parametric oscillator [\cite{stoler1970,stoler1974,ebarly1985,cervero1997}] can be written as ($\hbar=1$ herein) 

\beq
H = \omega a^{\dag} a +  \sum_{\jmath=1,2} \left( \frac{\Delta_{\jmath}}{2} \sigma_{\jmath}^{z} 
+  \lambda_{\jmath} \sigma_{\jmath}^{x} (a^{\dag} + a) \right)   + g ({a^{\dag}}^{2} + a^{2}),
\label{RPH}
\eeq
where the two nonidentical qubits are represented by the Pauli operators $(\sigma^{x}_{\jmath}, \sigma^{z}_{\jmath})$  having transition frequencies $\Delta_{\jmath}$. The single-mode quantum field is described by the annihilation and creation operators ($a, a^{\dagger}| \hat{n} \equiv a^{\dagger} a$) and the  frequency $\omega$. The coupling strength between  qubits and the field are denoted by $\lambda_{\jmath}$ and $g$ corresponds to the strength of the parametric oscillator. The Fock states 
$\{\hat{n} |n\rangle = n |n\rangle,\,n = 0, 1,\ldots;\;a \,|n\rangle = \sqrt{n}\,|n - 1\rangle, 
a^{\dagger}\, |n\rangle = \sqrt{n + 1}\,|n + 1\rangle\}$ provide the basis for the oscillator, whereas the eigenstates $\sigma^{x}_{1} |\pm 1\rangle \otimes \sigma^{x}_{2}  |\pm 1\rangle = \pm \,\ket{\pm 1} \otimes \pm \ket{\pm 1}$ span the space of the qubit. The Hamiltonian (\ref{RPH}) can be physically realized in the atom-photon interacting systems [\cite{czhu2020,agarwa2021}]. Various methods have been proposed to obtain the energy spectrum and eigenstates of the Rabi Hamiltonian, which are applicable to different parameter regimes. For example, we commonly use the well-known rotating wave approximation (RWA) [\cite{jaynes1963}] to probe the dynamical behaviour of the qubit-oscillator system for a weak coupling between the oscillator and the qubit having nearly identical frequencies. To investigate the regimes beyond the RWA, an adiabatic approximation scheme [\cite{irish2005,ashhab2010}] has been put forward in the far-off-resonance.

\par

To begin with the approximation, we  rewrite the Hamiltonian (\ref{RPH}) in terms of the delocalized qubit variables: $\{ S_{\pm}^{\mathcal{X}}=\frac{1}{2} (\sigma_{1}^{\mathcal{X}} \pm \sigma_{2}^{\mathcal{X}}), \; \mathcal{X} \in (x,y,z) \}$
\beq
H = H_{\mathcal{Q}} + \omega a^{\dag} a +  \sum_{\imath \in \pm}  
  \lambda_{\imath} S_{\imath}^{x} (a^{\dag} + a)    + g ({a^{\dag}}^{2} + a^{2}), \;
  H_{\mathcal{Q}} = \sum_{\imath \in \pm} \frac{\Delta_{\imath}}{2} S_{\imath}^{z},
\label{DH}
\eeq
where $\Delta_{\pm}=\Delta_{1} \pm \Delta_{2}$, and $\lambda_{\pm}=\lambda_{1} \pm \lambda_{2}$. Within the framework of  adiabatic approximation, we consider the qubit's energy splitting is  smaller compared to the oscillator's frequency i.e.  $\Delta_{\jmath} \ll \omega$. Now, the Hamiltonian for the oscillator degree of freedom can be obtained by posing  $\Delta_{\pm}=0$ in (\ref{DH}) and substituting the qubit variables with its eigenvalues: $\{\braket{\sigma^{x}_{\jmath}} = s_{\jmath} = \pm 1, \jmath \in (1,2) \}$. As a result, the delocalized qubit variables are replaced with their eigenvalues: $\braket{S_{\pm}^{x}} =(s_{1} \pm s_{2})/2$. Therefore, in the oscillator degree of freedom, the effective Hamiltonian is written as
\beq
H_{\mathcal{O}}=\omega a^{\dag}a + \lambda_{s_{1},s_{2}}    (a^{\dag} + a) + g ({a^{\dag}}^{2} + a^{2}), \; \lambda_{s_{1},s_{2}}  = \lambda_{1} s_{1} + \lambda_{2} s_{2}.
\label{EHO}
\eeq 
 The Hamiltonian $H_{\mathcal{O}}$ is  diagonalizable in the basis $\ket{n_{s_{1},s_{2}}}$ when $g = 0$. The  displaced number states read as: $\ket{n_{s_{1},s_{2}}}= \mathrm{D}^{\dagger}\left( \frac{\lambda_{s_{1},s_{2}}}{\omega}\right) \ket{n},
\, \mathrm{D}\left(\alpha \right) = \exp\left(\alpha a^{\dagger}- \alpha^{*}a \right),\, \alpha \in \mathbb{C}$, and the degenerate eigenenergies of $H_{\mathcal{O}}$  can be given as $E_{n} = 
\omega\big(n - \frac{\lambda^{2}_{s_{1},s_{2}}}{\omega^{2}}\big)$. The composite state of the system, consisting of displaced oscillator basis $\ket{n_{s_{1},s_{2}}}$  tensored with the two-qubit basis $\ket{s_{1},s_{2}}$, is used to block-diagonalize the full Hamiltonian, resulting in a non-degenerate energy eigen spectrum in the adiabatic approximation. To facilitate forthcoming calculations, we provide the formula to compute the overlap between the displaced number states [\cite{irish2005}]
\beq
\mathcal{M}_{m,n}(x) \equiv \braket{m|\mathrm{D}(x)|n} =
\begin{cases}
	x^{m-n}  \; \exp\big(-\frac{x^{2}}{2}\big) \;
	\sqrt{n!/m!} 
	\; L_n^{(m-n)}(x^{2}), & m \geq n \\
	(-x)^{n-m} \; \exp\big(-\frac{x^{2}}{2}\big) \;
	\sqrt{m!/n!} 
	\; L_m^{(n-m)}(x^{2}) & m < n,
	\label{mn}
\end{cases}
\eeq
where the associated Laguerre polynomial reads as  $L_{n}^{(j)}(\mathsf{x}) = \sum_{k = 0}^{n} 
\,(-1)^{k}\, \binom{n + j}{n - k}\,\frac{\mathsf{x}^{k}}{k!}$. The matrix element (\ref{mn}) leads to the identities: 
$ \mathcal{M}_{m,n}(x) =  (-1)^{n+m}  \mathcal{M}_{n,m}(x)$ and $\mathcal{M}_{m,n}(x)= (-1)^{n+m}\mathcal{M}_{m,n}(-x)$. In a similar way, when the parametric term  is present ($g \neq 0$), we use the Bogoliubov transformation [\cite{duan2019}] to diagonalize the Hamiltonian $H_{\mathcal{O}}$. This is equivalent to rewriting the Hamiltonian $H_{\mathcal{O}}$ in terms of the new operators  $(\widetilde{a},\widetilde{a}^{\dagger})$,  which follow the usual bosonic commutation relations.
\beq
H_{\mathcal{O}}= \Omega \, \widetilde{a}^{\dagger}\widetilde{a} - \frac{1}{2}(\omega-\Omega)
- \frac{\lambda^{2}_{s_{1},s_{2}}}{\omega+2g}, \quad \widetilde{a} = \mathrm{S}^{\dag}(r) \mathrm{D}^{\dag}(\eta_{s_{1},s_{2}}) a \mathrm{D}(\eta_{s_{1},s_{2}}) \mathrm{S}(r).
\label{BOH}
\eeq
The squeezing operator involved in (\ref{BOH}) reads as, $\mathrm{S}(\xi)=\exp((\xi {a^{\dag}}^{2}- \xi^{*} a^{2})/2)$, $\xi= r \exp(i \vartheta)$, $\xi \in \mathbb{C}$, and it satisfies the following unitary transformations:
\beq
\mathrm{S}^{\dag}(\xi) a \mathrm{S}(\xi) = \mu a + \nu a^{\dag}, \quad \mathrm{S}^{\dag}(\xi) a^{\dag} \mathrm{S}(\xi) = \mu a^{\dag} + \nu^{*} a, \quad \mu = \cosh(r), \; \nu = \exp(i \vartheta) \sinh(r),
\label{sqo}
\eeq
\begin{figure}
	\begin{center}
		\captionsetup[subfigure]{labelformat=empty}
		\subfloat[$(\mathsf{a})$]{\includegraphics[width=5.5cm,height=6cm]{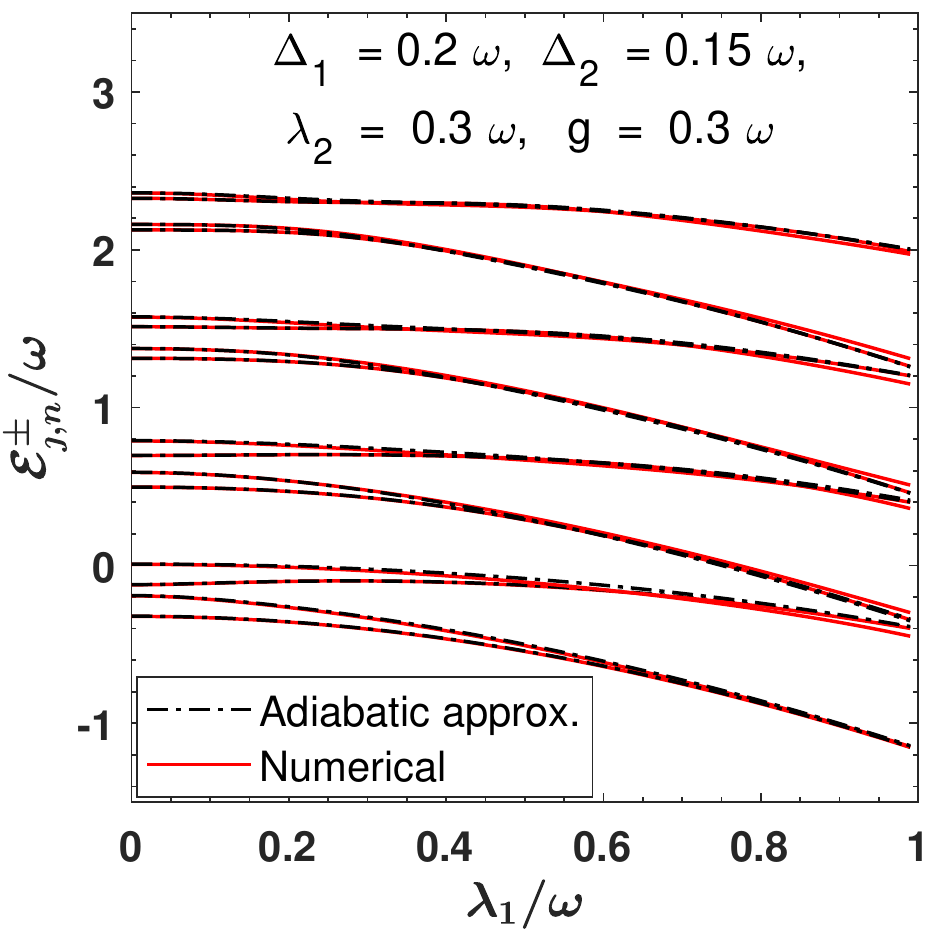}} 
		\captionsetup[subfigure]{labelformat=empty}
		\subfloat[$(\mathsf{b})$]{\includegraphics[width=5.5cm,height=6cm]{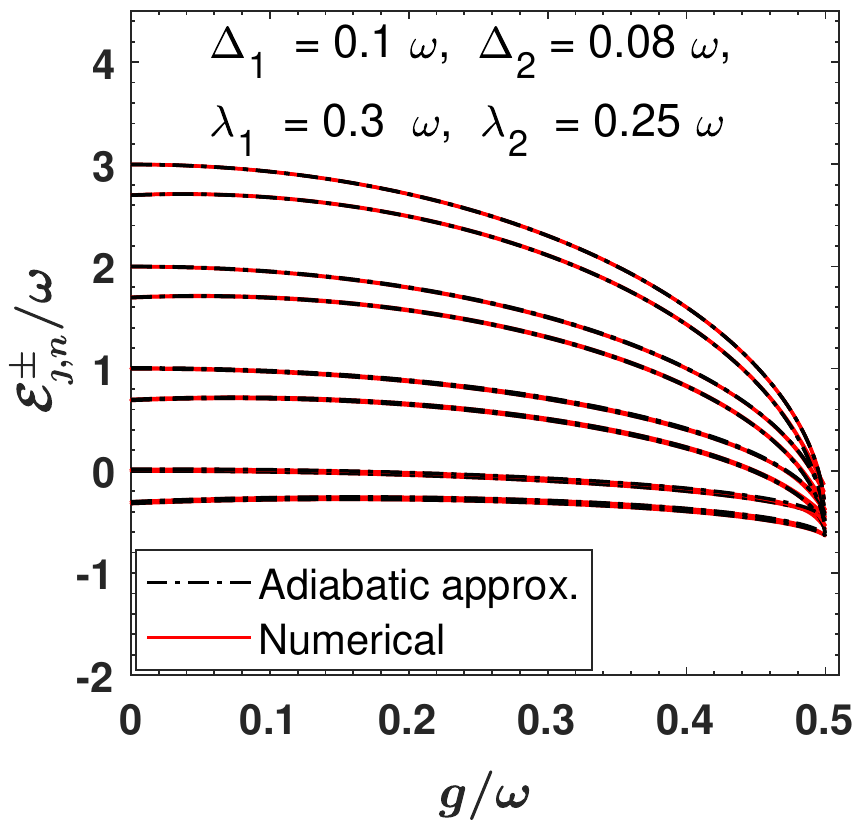}} 
		\caption{ Adiabatic approximation energy levels (\ref{ADEng}) (dotted-dashed) are compared with numerically-determined  energies (solid) in the far-off-resonance as a function of $(\mathsf{a})$  coupling strength $\lambda_1/\omega$,  and $(\mathsf{b})$ parametric strength $ g/\omega$. Our adiabatic approximation approach works well  far away from the resonance in the parameters regime as mentioned in the plots $(\mathsf{a})$ and $(\mathsf{b})$.}  
		\label{Fig_Eng_lam}
	\end{center}
\end{figure}
where $\Omega = \sqrt{\omega^{2}-4g^{2}}$, $r=arc \cosh \left( \sqrt{\frac{\omega + \Omega}{2 \Omega}} \right)$ and $\eta_{s_{1},s_{2}} = \sqrt{\frac{\omega + \Omega}{2 \Omega}}  \left(1+ \frac{\omega - \Omega}{2g} \right) \frac{\lambda_{s_{1},s_{2}}}{\omega + 2g} $. The effective Hamiltonian $H_{\mathcal{O}}$ (\ref{BOH}) can now be diagonalized in the following oscillator basis $\ket{r,n_{s_{1},s_{2}}}$
\beq
H_{\mathcal{O}} \ket{r,n_{s_{1},s_{2}}} = E_{n}^{\braket{s}} \ket{r,n_{s_{1},s_{2}}}, \quad
E_{n}^{s_{1},s_{2}} =(n+\frac{1}{2})\Omega - \frac{\omega}{2} - \frac{\lambda^{2}_{s_{1},s_{2}}}{\omega + 2g}, \quad
\ket{r,n_{s_{1},s_{2}}} = \mathrm{S}^{\dag}(r)\mathrm{D}^{\dag}( \eta_{s_{1},s_{2}}) \ket{n}.
\label{AOE}
\eeq
We note that at $g=0.5$, the eigenenergies of the $H_{\mathcal{O}}$ exhibit spectral collapse [\cite{ng1999,rico2020}]. This feature arises due to the presence of two-photon terms $(a^{2},a^{\dagger 2})$ in the Hamiltonian, and it can be understood by introducing the following phase-space variables: $x=  (1/\sqrt{2 \omega}) (a + a^{\dagger})$ and $p =  (\sqrt{\omega}/i\sqrt{2}) (a -  a^{\dagger})$, and re-writing the Hamiltonian (\ref{EHO}) as 
\beq
H_{\mathcal{O}} = \frac{p^{2}}{2\widetilde{m}} + \frac{1}{2} \widetilde{m} \Omega^{2} \left( x + \sqrt{\frac{2}{\omega}} \frac{\lambda_{s_{1},s_{2}}}{(2g + \omega)} \right)^{2} - \frac{\lambda_{s_{1},s_{2}}^{2}}{(2g+\omega)} - \frac{\omega}{2},
\label{EHO_phase}
\eeq
where $\widetilde{m} = \big (1 - \frac{2g}{\omega}\big)^{-1}$ and $\Omega  = \sqrt{\omega^{2} - 4g^{2}}$.  The position and energy shifted harmonic oscillator (\ref{EHO_phase}) has distinct spectrum for $\frac{2g}{\omega} < 1$. In the limit $\frac{2g}{\omega} \rightarrow 1$, the spectrum begins to collapse and becomes continuous with eigenfunctions that are Dirac-$ \delta $ normalizable, as the oscillator tends to behave like a free particle. Whereas for  $\frac{2g}{\omega} > 1 $,  we have an inverted oscillator whose spectrum remains continuous and the eigenfunctions are Dirac-$\delta$ normalizable [\cite{wolf2010}].

The Hamiltonian (\ref{DH}) is then  truncated into $4 \times 4$ blocks by using the oscillator basis (\ref{AOE})  tensored with the two-qubit basis: $\ket{s_{1},s_{2} ; r ,n_{_{s_{1},s_{2}}}}\equiv \ket{s_{1},s_{2}} \otimes \ket{ r ,n_{s_{1},s_{2}}}$,
\beq
\begin{pmatrix}
	E_{n}^{1,1} & \Delta_{1,n} & \Delta_{2,n} & 0   \\
	\Delta_{1,n}  & E_{n}^{-1,1} & 0 & \Delta_{2,n}\\
	\Delta_{2,n} & 0 & E_{n}^{1,-1}  & \Delta_{1,n}\\
	0 & \Delta_{2,n} &\Delta_{1,n}  & E_{n}^{-1,-1}
\end{pmatrix}, \; n\geq 0,
\label{4by4AD}
\eeq
where the off-diagonal terms are represented as
\bea
\Delta_{1,n} \!\!\! &=& \!\!\! \braket{-1,1;r,n_{-1,1}|H|1,1;r,n_{1,1}} =\frac{\Delta_{1}}{2}\mathcal{M}_{n,n}(-\zeta_{+})= \frac{\Delta_{1}}{2} \exp\left(- \frac{\zeta_{+}^{2}}{2}\right) L_{n}(\zeta_{+}^{2}), 
\nn \\
\Delta_{2,n} \!\!\! &=& \!\!\! \braket{1,-1;r,n_{1,-1}|H|1,1;r,n_{1,1}} =\frac{\Delta_{2}}{2}\mathcal{M}_{n,n}(-\zeta_{-})= \frac{\Delta_{2}}{2} \exp\left(- \frac{\zeta_{-}^{2}}{2}\right) L_{n}(\zeta_{-}^{2}),
\eea
and $\zeta_{\pm}=\eta_{1,1}-\eta_{\mp1,\pm1}$. The other off-diagonal elements of the Hamiltonian (\ref{4by4AD}) are followed from the reflection property of the displaced number states $\braket{n_{-1,1}|n_{1,1}}=\braket{n_{1,-1}|n_{-1,-1}}$, $\braket{n_{1,-1}|n_{1,1}}=\braket{n_{-1,1}|n_{-1,-1}}$. The diagonal elements can be written as $E_{n}^{1,1}=E_{n}^{-1,-1}$, and $E_{n}^{-1,1}=E_{n}^{1,-1}=E_{n}^{1,1}+2 \Lambda$ with $\Lambda=(2 \lambda_{1}\lambda_{2})/(\omega + 2g)$.
From the above matrix representation (\ref{4by4AD}), the adiabatic energies $(\mathcal{E}_{1,n}^{\pm},\mathcal{E}_{2,n}^{\pm})$ are obtained:
\beq
\mathcal{E}_{1,n}^{\pm}=E_{n}^{1,1}+\Lambda \pm \chi_{+}, \; 
\mathcal{E}_{2,n}^{\pm}=E_{n}^{1,1}+\Lambda \pm \chi_{-}, \quad  \chi_{\pm,n}=\sqrt{\Gamma_{\pm,n}^{2}+\Lambda^{2}},\;
\Gamma_{\pm,n} = \Delta_{1,n}\pm \Delta_{2,n}.
\label{ADEng}
\eeq
The corresponding adiabatic basis can be given as
\bea
\ket{\mathcal{E}_{1,n}^{\pm}} \!\!\! &=& \!\!\! 
\varepsilon_{\mp,n} \left( \ket{1,1;r,n_{1,1}} + \ket{-1,-1;r,n_{-1,-1}} \right)
\pm \frac{\Gamma_{+,n}}{|\Gamma_{+,n}|} \varepsilon_{\pm,n} \left( \ket{-1,1;r,n_{-1,1}} + \ket{1,-1;r,n_{1,-1}} \right), \quad \nn \\
\ket{\mathcal{E}_{2,n}^{\pm}} \!\!\! &=& \!\!\!
\kappa_{\mp,n}\left( \ket{1,1;r,n_{1,1}} - \ket{-1,-1;r,n_{-1,-1}} \right)
 \pm \frac{\Gamma_{-,n}}{|\Gamma_{-,n}|} \kappa_{\pm,n} \left(  \ket{-1,1;r,n_{-1,1}} - \ket{1,-1;r,n_{1,-1}} \right),
\label{ADstate}
\eea
here we abbreviate $\varepsilon_{\pm,n}=\frac{1}{2}\sqrt{\frac{\chi_{+,n} \pm \Lambda}{\chi_{+,n}}}$ and $\kappa_{\pm,n}=\frac{1}{2} \sqrt{\frac{\chi_{-,n}\pm \Lambda}{\chi_{-,n}}}$.
The completeness relation of the orthonormal basis 
(\ref{ADstate}) now reads as: 
\beq
 \sum_{\imath \in \pm}\sum_{n = 0}^{\infty} 
\left(| \mathcal{E}_{1,n}^{\imath} \rangle \langle  \mathcal{E}_{1,n}^{\imath}| +  | \mathcal{E}_{2,n}^{\imath} \rangle \langle  \mathcal{E}_{2,n}^{\imath}|\right) = \mathbb{I}.
\label{completeness}
\eeq 
 
\par

\section{Time evolution of the reduced density matrices}
\setcounter{equation}{0}
\label{dmatrix}
After completing the above construction of the energy eigenstates $(\ref{ADstate})$, we investigate the impact of parameter $ g $ on two-qubit subsystem, specifically the relative entropy of coherence, geometric discord and concurrence in details. The initial state of the composite system reads as: $\ket{\psi(0)} = (\cos\theta\ket{1,1} + \exp(i\phi) \sin \theta \ket{-1,-1})  \otimes \ket{\alpha}$, where $\ket{\alpha}$ is the coherent state of the oscillator. The time evolution of the initial state is
\beq 
\ket{\psi(t)} =\sum_{\imath \in \pm}\;\sum_{\jmath =1,2}\;\sum_{n = 0}^{\infty} 
 \mathcal{C}_{\jmath,n}^{\imath}(t)| \mathcal{E}_{\jmath,n}^{\imath} \rangle  , \quad \mathcal{C}_{\jmath,n}^{\imath}(t) = \mathcal{C}_{\jmath,n}^{\imath} \exp(-i \mathcal{E}_{\jmath,n}^{\imath} t),
\label{wave_time}
\eeq
where the coefficients read as:

\bea
\mathcal{C}_{1,n}^{\pm} \!\!\! &=& \!\!\! 
 \varepsilon_{\mp,n} \left( \cos\theta \; \braket{r,n_{1,1}|\alpha}
+ \; \exp(i \phi) \sin \theta \braket{r,n_{-1,-1}|\alpha}
\right), \nn \\
\mathcal{C}_{2,n}^{\pm} \!\!\! &=& \!\!\! 
\kappa_{\mp,n} \left( \cos\theta \; \braket{r,n_{1,1}|\alpha}
- \; \exp(i \phi) \sin \theta \braket{r,n_{-1,-1}|\alpha}
\right), 
\label{coeff}
\eea
and,
\bea
\braket{r,n_{\pm1,\pm1}|\alpha} &=& \frac{1}{\sqrt{\mu\, n!}}  \left(i\sqrt{\frac{\nu}{2\mu}} \right)^{n} 
 \exp \left(-(\mu -\nu) \frac{\eta_{\pm1,\pm1}^{2}}{2\mu} - \frac{|\alpha|^{2}}{2}-\frac{\alpha^{2} \nu}{2\mu}-\frac{\eta_{\pm1,\pm1}\alpha}{\mu} \right) \nn \\ 
&& \times \; \mathrm{H}_{n}\left(- \frac{i((\mu-\nu)\eta_{\pm1,\pm1}+\alpha)}{\sqrt{2 \mu \nu}} \right).
\label{inner_product}
\eea
The coefficients (\ref{coeff}) are calculated using the inner product relationships given below:
\beq
\mathrm{S}(\xi) \mathrm{D}(\alpha) \ket{0} = \exp \left( -\frac{|\alpha|^{2}}{2} - \frac{\alpha^{2} \nu^{*}}{2 \mu}\right)
\sum_{n=0}^{\infty} \frac{i^{n}}{\sqrt{n! \mu}} \left( \frac{\nu}{2 \mu}\right)^{\!\! \frac{n}{2}}
\mathrm{H}_{n}\left( \frac{-i\alpha}{\sqrt{2 \mu \nu}} \right) \ket{n}, \, 
 \mathrm{D}(\alpha) \mathrm{S}(\xi) = \mathrm{S}(\xi) \mathrm{D}(\alpha \mu - \alpha^{*} \nu).
\label{def_sq}
\eeq
\bea
\braket{\beta|\xi,n_{\pm}} = \frac{1}{\sqrt{\mu n!}} \left(-i \sqrt{\frac{\nu^{*}}{2\mu}} \right)^{n}
\exp \left(- \frac{|\alpha|^{2}}{2} + \frac{\alpha^{2}\nu^{*}}{2\mu} 
- \frac{|\beta|^{2}}{2} - \frac{\beta^{*2}\nu}{2\mu} \mp \frac{\alpha \beta^{*}}{\mu}\right)
\mathrm{H}_{n}\left( \frac{i (\pm \mu \alpha^{*}\mp \alpha \nu^{*}+ \beta^{*})}{\sqrt{2 \mu \nu^{*}}}\right), \quad
\label{def_cinner}
\eea
with the following definition $\ket{\xi,n_{\pm}}=S^{\dagger}(\xi)  D^{\dagger}(\pm \alpha)\ket{n}$, for the parameters $\xi, \alpha \in \mathbb{C}$.  The inner products (\ref{def_cinner}) can be obtained using the expressions mentioned in (\ref{def_sq}).  Here, the Hermite polynomials are given by the exponential generating function [\cite{gradshteyn2007}]: $\exp(2 \,\mathsf{x} \mathsf{t}-\mathsf{t}^{2}) = \sum_{n=0}^{\infty} \frac{\mathrm{H}(\mathsf{x}) \mathsf{t}^{n}}{n!}$. 
The following identity [\cite{andrews1999}] should be used to demonstrate the normalization of the state $\ket{\psi(t)}$: $\braket{\psi(t)|\psi(t)}=1$,
\beq
\sum_{n=0}^{\infty}\dfrac{\mathsf{t}^{n}}{2^{n}n!} \mathrm{H}_{n}(\mathsf{x}) \mathrm{H}_{n}(\mathsf{y}) =
\dfrac{1}{\sqrt{1-\mathsf{t}^{2}}}\,
\exp\left(-\dfrac{(\mathsf{tx})^{2}-2\mathsf{txy}+
	(\mathsf{ty})^{2}}{ 1-\mathsf{t}^{2}} \right).
\label{Hermite_id} 
\eeq
Thereafter, the time evolution of the density matrix of the total system can be represented as $\rho_{_{\text{tot}}}(t) \equiv \ket{\psi(t)} \bra{\psi(t)}$. By partial tracing over the oscillator-Hilbert space, one can obtain the reduced density matrix for the two-qubit subsystem as
\beq
\varrho(t) \equiv \mathrm{Tr}_{\mathcal{O}} \rho_{_{\text{tot}}}(t) = \sum_{s_{1},s_{2} \atop \in \{\pm 1\}} \sum_{s'_{1},s'_{2}\atop \in \{\pm 1\}} \varrho_{_{s_{1},s_{2};s'_{1},s'_{2}}}(t) \ket{s_{1},s_{2}} \bra{s'_{1},s'_{2}}.
\label{rdens}
\eeq
We will exclude the explicit time dependence from the matrix elements of $\varrho(t)$ in the future for notational simplicity. Its  diagonal elements are explicitly constructed as follows:
\bea
\varrho_{_{1,1;1,1}} \!\!\!\!\! &=& \!\!\!\!\! \sum_{n=0}^{\infty} \Big ( \mathcal{A}_{n}^{+}(t)  + 2 \, \mathrm{Re} 
\Big \lgroup \varepsilon_{-,n} \mathcal{C}_{1,n}^{+}(t)
\Big( \widetilde{\mathcal{C}}_{2,n}^{+}(t)^{*}  + \varepsilon_{+,n}
\mathcal{C}_{1,n}^{-}(t)^{*}  
\Big) + \, \varepsilon_{+,n} \kappa_{+,n} \mathcal{C}_{1,n}^{-}(t) \, \mathcal{C}_{2,n}^{-}(t)^{*} \qquad \nn \\
&& \!\!\!   + \; \kappa_{-,n} \mathcal{C}_{2,n}^{+}(t)  \Big( \varepsilon_{+,n}
\mathcal{C}_{1,n}^{-}(t)^{*}  + \kappa_{+,n} \mathcal{C}_{2,n}^{-}(t)^{*}
\Big) \Big \rgroup \Big ), \nn \\
\varrho_{_{-1,1;-1,1}} \!\!\!\!\! &=& \!\!\!\!\! \sum_{n=0}^{\infty} \Big ( \mathcal{A}_{n}^{-}(t)  - 2 \, \mathrm{Re} 
\Big \lgroup \varepsilon_{+,n} \varepsilon_{-,n} \mathcal{C}_{1,n}^{+}(t) \mathcal{C}_{1,n}^{-}(t)^{*} +
\kappa_{+,n}\kappa_{-,n} \mathcal{C}_{2,n}^{+}(t) \mathcal{C}_{2,n}^{-}(t)^{*}  \qquad \nn \\
&& \!\!\!   - \;
\frac{\Gamma_{+,n}}{|\Gamma_{+,n}|} \frac{\Gamma_{-,n}}{|\Gamma_{-,n}|} 
\, \widetilde{\mathcal{C}}_{1,n}^{-}(t) \, \widetilde{\mathcal{C}}_{2,n}^{-}(t)^{*} \Big \rgroup \Big), \nn \\
\varrho_{_{1,-1;1,-1}} \!\!\!\!\! &=& \!\!\!\!\!  \sum_{n=0}^{\infty} \Big ( \mathcal{A}_{n}^{-}(t)  - 2 \, \mathrm{Re} 
\Big \lgroup \varepsilon_{+,n} \varepsilon_{-,n} \mathcal{C}_{1,n}^{+}(t) \mathcal{C}_{1,n}^{-}(t)^{*} +
\kappa_{+,n}\kappa_{-,n} \mathcal{C}_{2,n}^{+}(t) \mathcal{C}_{2,n}^{-}(t)^{*}  \qquad \nn \\
&& \!\!\!   + \;
\frac{\Gamma_{+,n}}{|\Gamma_{+,n}|} \frac{\Gamma_{-,n}}{|\Gamma_{-,n}|} 
\, \widetilde{\mathcal{C}}_{1,n}^{-}(t) \, \widetilde{\mathcal{C}}_{2,n}^{-}(t)^{*} \Big \rgroup \Big), \nn \\
\varrho_{_{-1,-1;-1,-1}} \!\!\!\!\! &=& \!\!\!\!\! \sum_{n=0}^{\infty} \Big ( \mathcal{A}_{n}^{+}(t)  - 2 \, \mathrm{Re} 
\Big \lgroup \varepsilon_{-,n} \mathcal{C}_{1,n}^{+}(t)
\Big( \widetilde{\mathcal{C}}_{2,n}^{+}(t)^{*}  - \varepsilon_{+,n}
\mathcal{C}_{1,n}^{-}(t)^{*}  
\Big) + \, \varepsilon_{+,n} \kappa_{+,n} \mathcal{C}_{1,n}^{-}(t) \, \mathcal{C}_{2,n}^{-}(t)^{*} \qquad \nn \\
&& \!\!\!   + \; \kappa_{-,n} \mathcal{C}_{2,n}^{+}(t)  \Big( \varepsilon_{+,n}
\mathcal{C}_{1,n}^{-}(t)^{*}  - \kappa_{+,n} \mathcal{C}_{2,n}^{-}(t)^{*}
\Big) \Big \rgroup \Big ), 
\label{rho_diag}
\eea
where,
\bea
\mathcal{A}_{n}^{\pm}(t) \!\!\! &=& \!\!\! \varepsilon_{\mp,n}^{2} |\mathcal{C}_{1,n}^{+}(t)|^{2} 
+ \kappa_{\mp,n}^{2} |\mathcal{C}_{2,n}^{+}(t)|^{2} 
+ \varepsilon_{\pm,n}^{2} |\mathcal{C}_{1,n}^{-}(t)|^{2} 
+ \kappa_{\pm,n}^{2} |\mathcal{C}_{2,n}^{-}(t)|^{2}, \nn \\
\widetilde{\mathcal{C}}_{1,n}^{\pm}(t)\!\!\! &=& \!\!\!
\varepsilon_{\mp,n} \mathcal{C}_{1,n}^{+}(t) \pm 
\varepsilon_{\pm,n} \mathcal{C}_{1,n}^{-}(t), \quad 
\widetilde{\mathcal{C}}_{2,n}^{\pm}(t)  = 
\kappa_{\mp,n} \mathcal{C}_{2,n}^{+}(t) \pm 
\kappa_{\pm,n} \mathcal{C}_{2,n}^{-}(t).
\eea
The trace i.e. the sum of diagonal elements of the reduced density matrix for the two-qubit  (\ref{rho_diag})  is preserved:  $\mathrm{Tr}\varrho(t)=1$. The off-diagonal elements which reflect the Hermiticity property of $\varrho(t)$ are constructed as
\bea
\varrho_{_{1,1;-1,1}}  \!\!\!\!\! &=& \!\!\!\!\! \sum_{n,m=0}^{\infty}
\left( \widetilde{\mathcal{C}}_{1,n}^{+}(t) +
\widetilde{\mathcal{C}}_{2,n}^{+}(t) \right)
\Big( \frac{\Gamma_{+,m}}{|\Gamma_{+,m}|} \; \widetilde{\mathcal{C}}_{1,m}^{-}(t)^{*} +
\frac{\Gamma_{-,m}}{|\Gamma_{-,m}|} \; \widetilde{\mathcal{C}}_{2,m}^{-}(t)^{*} \Big)
 \mathcal{M}_{m,n}(-\zeta_{+}),  \nn \\ 
\varrho_{_{1,1;1,-1}}  \!\!\!\!\! &=& \!\!\!\!\! \sum_{n,m=0}^{\infty}
\left( \widetilde{\mathcal{C}}_{1,n}^{+}(t) +
\widetilde{\mathcal{C}}_{2,n}^{+}(t) \right)
\Big( \frac{\Gamma_{+,m}}{|\Gamma_{+,m}|} \; \widetilde{\mathcal{C}}_{1,m}^{-}(t)^{*} -
\frac{\Gamma_{-,m}}{|\Gamma_{-,m}|} \; \widetilde{\mathcal{C}}_{2,m}^{-}(t)^{*} \Big)
\mathcal{M}_{m,n}(-\zeta_{-}),  \nn \\ 
\varrho_{_{1,1;-1,-1}}  \!\!\!\!\! &=& \!\!\!\!\! \sum_{n,m=0}^{\infty}
\left( \widetilde{\mathcal{C}}_{1,n}^{+}(t) +
\widetilde{\mathcal{C}}_{2,n}^{+}(t) \right)
\left( \widetilde{\mathcal{C}}_{1,m}^{+}(t)^{*} -
\widetilde{\mathcal{C}}_{2,m}^{+}(t)^{*} \right) \mathcal{M}_{m,n}(-2\eta_{1,1}), \nn
\eea 
\bea
\varrho_{_{-1,1;1,-1}}  \!\!\!\!\! &=& \!\!\!\!\! \sum_{n,m=0}^{\infty}  \Big( \frac{\Gamma_{+,n}}{|\Gamma_{+,n}|} 
\widetilde{\mathcal{C}}_{1,n}^{-}(t) + \frac{\Gamma_{-,n}}{|\Gamma_{-,n}|} \widetilde{\mathcal{C}}_{2,n}^{-}(t) \Big) \, \Big ( 
\frac{\Gamma_{+,m}}{|\Gamma_{+,m}|} \, 
\widetilde{\mathcal{C}}_{1,m}^{-}(t)^{*} - 
\frac{\Gamma_{-,m}}{|\Gamma_{-,m}|} \, 
\widetilde{\mathcal{C}}_{2,m}^{-}(t)^{*} \Big) 
\mathcal{M}_{m,n}(2\eta_{1,-1}), \quad \nn \\
\varrho_{_{-1,1;-1,-1}}  \!\!\!\!\! &=& \!\!\!\!\! \sum_{n,m=0}^{\infty}
\Big( \frac{\Gamma_{+,n}}{|\Gamma_{+,n}|} \; \widetilde{\mathcal{C}}_{1,n}^{-}(t) +
\frac{\Gamma_{-,n}}{|\Gamma_{-,n}|} \; \widetilde{\mathcal{C}}_{2,n}^{-}(t) \Big)
\left( \widetilde{\mathcal{C}}_{1,m}^{+}(t)^{*} -
\widetilde{\mathcal{C}}_{2,m}^{+}(t)^{*} \right)
\mathcal{M}_{m,n}(-\zeta_{-}),  \nn \\ 
\varrho_{_{1,-1;-1,-1}}  \!\!\!\!\! &=& \!\!\!\!\! \sum_{n,m=0}^{\infty}
\Big( \frac{\Gamma_{+,n}}{|\Gamma_{+,n}|} \; \widetilde{\mathcal{C}}_{1,n}^{-}(t) -
\frac{\Gamma_{-,n}}{|\Gamma_{-,n}|} \; \widetilde{\mathcal{C}}_{2,n}^{-}(t) \Big)
\left( \widetilde{\mathcal{C}}_{1,m}^{+}(t)^{*} -
\widetilde{\mathcal{C}}_{2,m}^{+}(t)^{*} \right)
\mathcal{M}_{m,n}(-\zeta_{+}). 
\label{rho_offdiag}
\eea
In the evaluation of the off-diagonal elements of the reduced density matrix (\ref{rho_offdiag}), we use the following  inner products of the displaced number states:
\bea
\braket{m_{1,1}|n_{-1,-1}} \!\!\!\!\! &=&  \!\!\!\!\! (-1)^{n+m} \braket{m_{-1,-1}|n_{1,1}} = \mathcal{M}_{m,n}(2\eta_{1,1}),   \nn \\
\braket{m_{1,-1}|n_{-1,1}} \!\!\!\!\! &=&  \!\!\!\!\! (-1)^{n+m} \braket{m_{-1,1}|n_{1,-1}} = \mathcal{M}_{m,n}(2\eta_{1,-1}), \nn \\
\braket{m_{1,1}|n_{\mp1,\pm1}} \!\!\!\!\! &=&  \!\!\!\!\! (-1)^{n+m} \braket{m_{\mp1,\pm1}|n_{1,1}} = \mathcal{M}_{m,n}(\zeta_{\pm}), \nn \\
\braket{m_{\pm1,\mp1}|n_{-1,-1}} \!\!\!\!\! &=&  \!\!\!\!\! (-1)^{n+m} \braket{m_{-1,-1}|n_{\pm1,\mp1}} = \mathcal{M}_{m,n}(\zeta_{\pm}). 
\label{prop_dns}
\eea

\par
The reduced density matrices for the individual qubit subsystems are provided below:
\bea
\varrho_{_{\mathcal{Q}_{1}}} \equiv \mathrm{Tr}_{\mathcal{Q}_{2}} \varrho(t) &= &
\begin{pmatrix}
\varrho_{_{1,1;1,1}} + \varrho_{_{-1,1;-1,1}}  &  \varrho_{_{1,1;1,-1}} + \varrho_{_{-1,1;-1,-1}}\\
\varrho_{_{1,1;1,-1}}^{*} + \varrho_{_{-1,1;-1,-1}}^{*}  &  \varrho_{_{1,-1;1,-1}} + \varrho_{_{-1,-1;-1,-1}}
\end{pmatrix}, \\
\varrho_{_{\mathcal{Q}_{2}}} \equiv \mathrm{Tr}_{\mathcal{Q}_{1}} \varrho(t) &= &
\begin{pmatrix}
\varrho_{_{1,1;1,1}} + \varrho_{_{1,-1;1,-1}}  &  \varrho_{_{1,1;-1,1}} + \varrho_{_{1,-1;-1,-1}}\\
\varrho_{_{1,1;-1,1}}^{*} + \varrho_{_{1,-1;-1,-1}}^{*}  &  \varrho_{_{-1,1;-1,1}} + \varrho_{_{-1,-1;-1,-1}}
\end{pmatrix}.
\eea
\par
Similarly, partial tracing over any one of the two qubits from the density matrix of the composite system, yields the reduced density matrix consisting of single qubit and the oscillator 
\beq
\varpi(t) = \mathrm{Tr}_{\mathcal{Q}_{2}}\rho_{_{\text{tot}}}(t).
\label{rho_qubit-osc} 
\eeq
Here, $\mathrm{Tr}_{\mathcal{Q}_{2}}$  denotes a partial trace performed over the second qubit degree of freedom, and $\varpi(t)$ obeys the normalization property: $\mathrm{Tr}(\varpi(t))=1$. The explicit form of $\varpi(t)$ is given in the Appendix \ref{appendix:Ap1}.  Therefore,  the oscillator's  reduced density matrix  can be obtained 
\beq
\rho_{\mathcal{O}}(t)= \mathrm{Tr}_{\mathcal{Q}_{1}} \varpi(t),
\label{osc_den}
\eeq
\bea
\rho_{\mathcal{O}}(t)\!\!\!\!&=&\!\!\!\! \sum_{n,m=0}^{\infty} \left( \mathcal{F}_{n,m}^{(1)} \ket{r,n_{1,1}} \bra{r,m_{1,1}}
+ \mathcal{F}_{n,m}^{(-1)} \ket{r,n_{-1,-1}} \bra{r,m_{-1,-1}}
\right. \quad \quad \nn \\
& & +  \left. \mathcal{F}_{n,m}^{(2)} \ket{r,n_{-1,1}} \bra{r,m_{-1,1}} +\mathcal{F}_{n,m}^{(-2)} \ket{r,n_{1,-1}} \bra{r,m_{1,-1}}\right). \quad \quad
\label{osc_den}
\eea
Now, the Von Neumann entropy for the two-qubit subsystem is defined as $\mathrm{S}(\varrho) \equiv  - \hbox{Tr} (\varrho \log_{2} \varrho)$. Since our total system is in a pure state,  the entropy of the oscillator degree of freedom is equal to the entropy of the two-qubit subsystem [\cite{araki2002}] i.e. $  S(\rho_{\mathcal{O}}) =  S(\varrho) $.
\begin{figure}
	\begin{center}
	    \captionsetup[subfigure]{labelformat=empty}
		\subfloat[$(\mathsf{a})$]{\includegraphics[width=4cm,height=3.5cm]{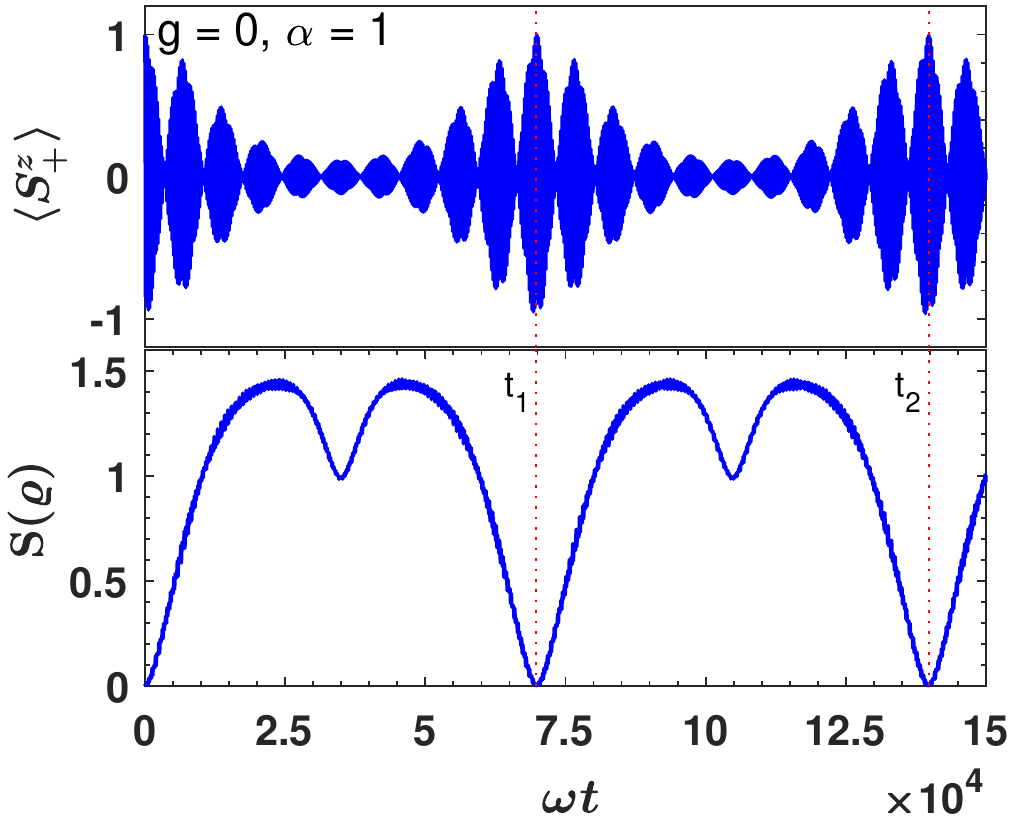}} 
		\captionsetup[subfigure]{labelformat=empty}
		\subfloat[$(\mathsf{b})$]{\includegraphics[width=4cm,height=3.5cm]{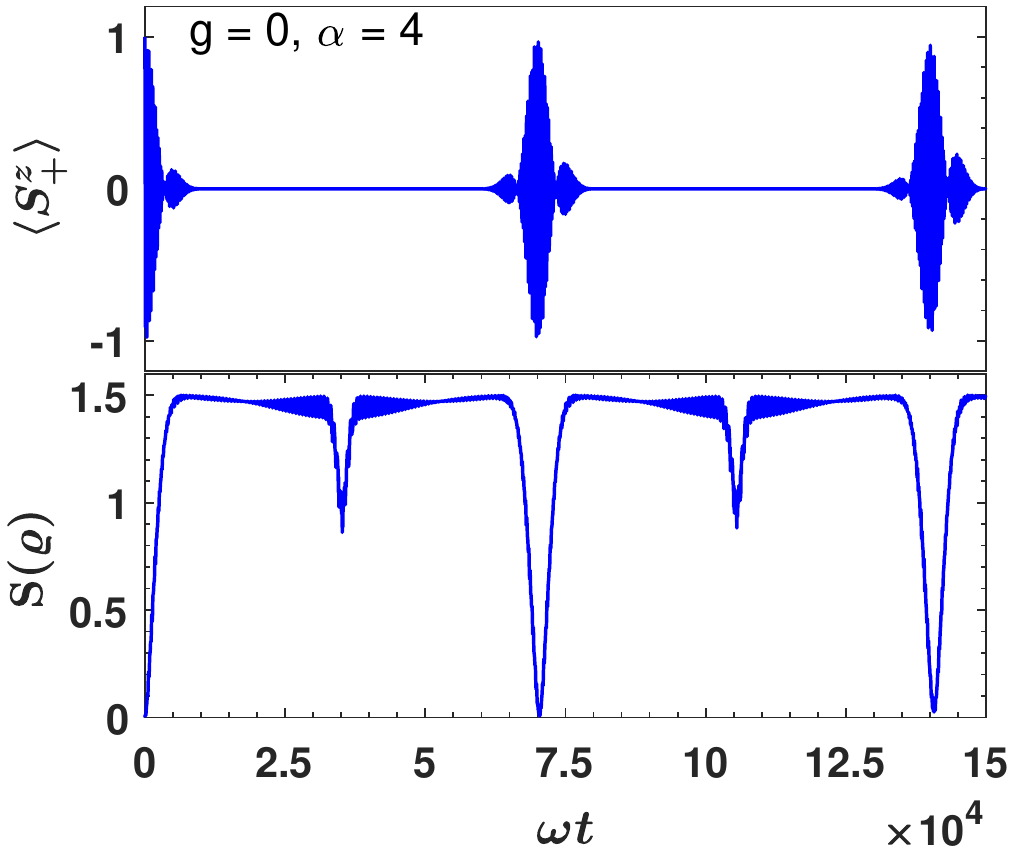}}
		\captionsetup[subfigure]{labelformat=empty}
		\subfloat[$(\mathsf{c})$]{\includegraphics[width=4cm,height=3.5cm]{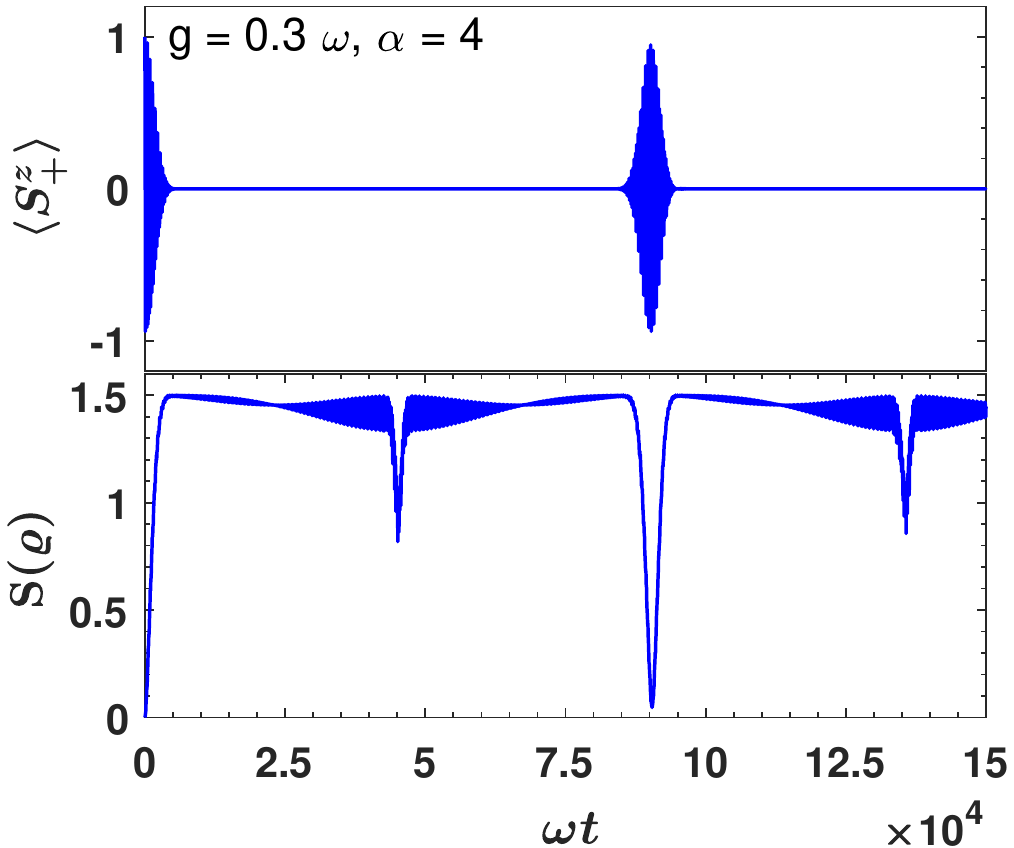}} 
		\captionsetup[subfigure]{labelformat=empty}
		\subfloat[$(\mathsf{d})$]{\includegraphics[width=4cm,height=3.5cm]{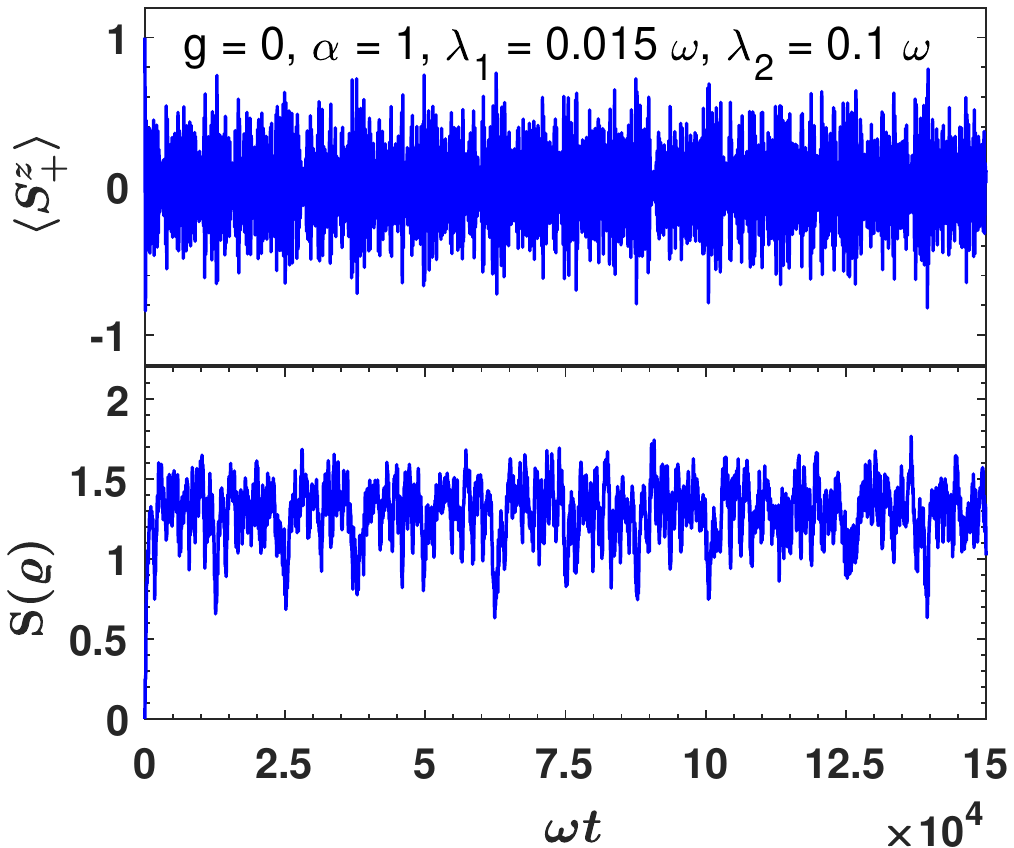}} 
		\captionsetup[subfigure]{labelformat=empty}
		\caption{The time evolution of the two-qubit population inversion (\ref{qubit_inv}) and the Von Neumann entropy $( \mathrm{S}(\varrho))$ are shown. Here, all the plots  are presented for the initially factorized two-qubit 
	    state  $(\theta = 0, \; \phi  = 0)$  by considering  a simple case where  the qubit's frequencies $(\Delta_{1},  \Delta_{2})$ are same and equal to $ 0.1 \; \omega$. The coupling strenghts  $( \lambda_{1}, \lambda_{2})$ in  $(\mathsf{a})$,  $(\mathsf{b})$ and   $(\mathsf{c})$ are also same and it equals $ 0.015 \; \omega$, whereas the plot $(\mathsf{d})$ is depicted for unequal coupling strength.  The two vertical dotted lines in $(\mathsf{a})$, correspond to the revivals  at the scaled times  $ t_{1} = 6.975 \times 10^{4}$ and $ t_{2} = 13.980 \times 10^{4}$. The values of $\mathrm{S}(\varrho)$ at these two revival times $(t_{1},t_{2})$ are $(0.0008, 0.0003)$. }  
		\label{Revival}
	\end{center}
\end{figure}
\par We will investigate the effect of parametric oscillator $(g)$ on the various quantities associated with the two-qubit subsystem as well as its individual qubit subsytems using this explicit description of the reduced density matrices.

\section{Revival and collapse}
\label{Rev_col}
With the above construction of the  reduced density matrix for the two-qubit subsystem (\ref{rdens}), we will study the revival and collapse phenomenon arising in the two-qubit population inversion and its explicit dependence on the parametric oscillator. The expression of the two-qubit population inversion in this case can be written as 
\bea
\langle S_{+}^{z} \rangle \equiv \frac{1}{2}\left( \langle \sigma_{1}^{z}\rangle +\langle \sigma_{2}^{z}\rangle \right) \!\!\!\! &=& \!\!\!\! \varrho_{_{1,1;1,1}}  -  \varrho_{_{-1,-1;-1,-1}}.
\label{qubit_inv} 
\eea
The physical reason for the occurrence of the revival-collapse phenomenon  is due to the periodic exchange of  energy between the qubits and oscillator mode. It is observed that for $\zeta_{\pm}^{2}\ll 1$,  the two-qubit population inversion exhibits the familiar  revival and collapse particularly revivals with echoes (Fig. \ref{Revival} $(\mathsf{a})$) [\cite{mvs,buch2000}]. This is because the phases of different eigenstates  in the superposition state  evolve with different frequencies in time which causes the decay of the oscillation. During the evolution, a new superposition state is formed after the revival time  $t_{R}$. The eigenstates which are in phase with each other, contribute strongly to the new superposition  state. However, the eigenstates that are out of phase actually lead to dephasing and give rise to the phenomena of echo. Moreover, with  increasing  $\alpha$,  the incoherent evolution of the superposition state occurs which causes the randomizations in phases of the eigenstates, and subsequently the amplitude of echo decreases (Fig. \ref{Revival} $(\mathsf{b})$).
\par
On the other hand, any change in the  $\zeta_{\pm}^{2}$, causes a change in the population inversion. Therefore, increasing the $\zeta_{\pm}^{2}$ leads to elongation of the revival time while major revivals in the population inversion maintaining almost same amplitude which is evident from Figs. \ref{Revival} $(\mathsf{b})$ and $(\mathsf{c})$.  Furthermore, considering the Von Neumann entropy, it is apparent that when revivals occur in the  population inversion, the state corresponding to the two-qubit subsystem will be a nearly pure state. It is noticed that with increasing $\zeta_{\pm}^{2}$, the number of isolated revivals also diminish over the same time period. The periodic behaviour of the  population inversion is completely lost as coupling strength increases (Fig. \ref{Revival} $(\mathsf{d})$).  This is due to the fact that as the coupling strength increases, a large number of incommensurate frequencies begin to participate, causing the dynamics to be erratic. As a consequence, the energy exchange between the qubits and the oscillator is no longer periodic.
\par
We estimate the revival time under the condition $\zeta_{\pm}^{2}\ll1$. Firstly, the energy levels in (\ref{ADEng}) are approximated by keeping  the  terms  in the Laguerre polynomials up to the  $O(\zeta_{\pm}^{2})$. After some calculations these are  explicitly written as
\bea
\mathcal{E}_{1,n}^{\pm}  \!\!\!& \approx &\!\!\! E_{n}^{1,1} + \Lambda \pm \left( \Delta_{1,0} + \Delta_{2,0} + \Lambda-\frac{\Delta_{1,0} + \Delta_{2,0}}{\Delta_{1,0} + \Delta_{2,0}+\Lambda} \left (1+ (  \zeta_{+}^{2} \Delta_{1,0} + \zeta_{-}^{2} \Delta_{2,0}) n \right) \right), \quad \nn  \\
\mathcal{E}_{2,n}^{\pm} \!\!\! & \approx & \!\!\! E_{n}^{1,1} + \Lambda \pm \left( \Delta_{1,0} - \Delta_{2,0} + \Lambda-\frac{\Delta_{1,0} - \Delta_{2,0}}{\Delta_{1,0} - \Delta_{2,0} + \Lambda} \left (1+ (\zeta_{+}^{2} \Delta_{1,0}  - \zeta_{-}^{2} \Delta_{2,0}) n \right ) \right).
\label{revival_time}
\eea
Let us consider a simple case for the Fig. \ref{Revival} $(\mathsf{a})$, where the frequencies of the two qubits are assumed to be same, say  $\Delta $,  and the coupling constants are equal $\lambda$,  which leads to the  $\zeta_{\pm} = 2 \eta_{1,0}$. Therefore, the reduced form of (\ref{revival_time}) reads as
\bea
\mathcal{E}_{1,n}^{\pm}  =  E_{n}^{1,1} +  \Lambda \pm \bigg( \widetilde{\Delta} + \Lambda -\frac{\widetilde{\Delta}}{\widetilde{\Delta} + \Lambda} \big (1+ 4 \eta_{1,0}^{2} \widetilde{\Delta}  n\big) \bigg), \quad  \mathcal{E}_{2,n}^{\pm} =  E_{n}^{1,1} + \Lambda \pm \Lambda, \quad \widetilde{\Delta}  = \Delta \exp {(- 2 \eta_{1,0}^{2})}.
\label{modified_eng}  
\eea
\par
In the study of Rabi oscillations, the expression of two-qubit population inversion $(\langle S_{+}^{z} \rangle) $   is useful for estimating the time period of revival. We utilize the Eq. (\ref{modified_eng}) to approximate the time-dependent phase factors associated with  the population inversion. The dominant frequency appearing from phase factors that are proportional to $\zeta_{\pm}^{2}$  and linear in $n$ are used to calculate the revival time. Following this, the order of time period of revival observed in the population inversion can be estimated as $O\left(2\pi/( (2\eta_{1,0})^{2}\widetilde{\Delta})\right)$. The successive revival times given up to a proportionality constant are  $ t_{R} \sim  2 \pi k/ ( (2 \eta_{1,0})^{2} \widetilde{\Delta}), \; k = 0, \; 1, \; 2,...$ For example, in the Fig. \ref{Revival} $(\mathsf{a})$, the proportionality constant for the estimated revival times at $t_{1}$ and $t_{2}$ are $0.999$ and $1.001$ respectively. The discrepancy in revival times is observed to be less than $1\%$.
\begin{figure}
	\begin{center}
		\captionsetup[subfigure]{labelformat=empty}
		\subfloat[$(\mathsf{a})$]{\includegraphics[width=4.15cm,height=3.8cm]{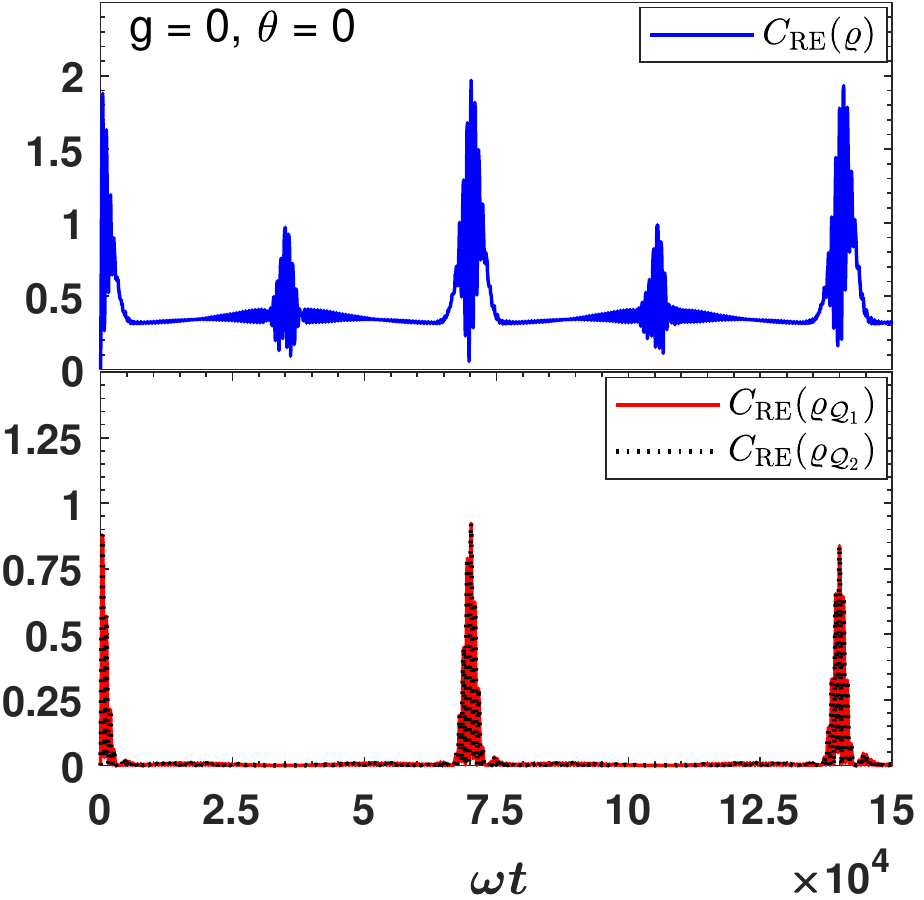}} 
		\captionsetup[subfigure]{labelformat=empty}
		\subfloat[$(\mathsf{b})$]{\includegraphics[width=4.15cm,height=3.8cm]{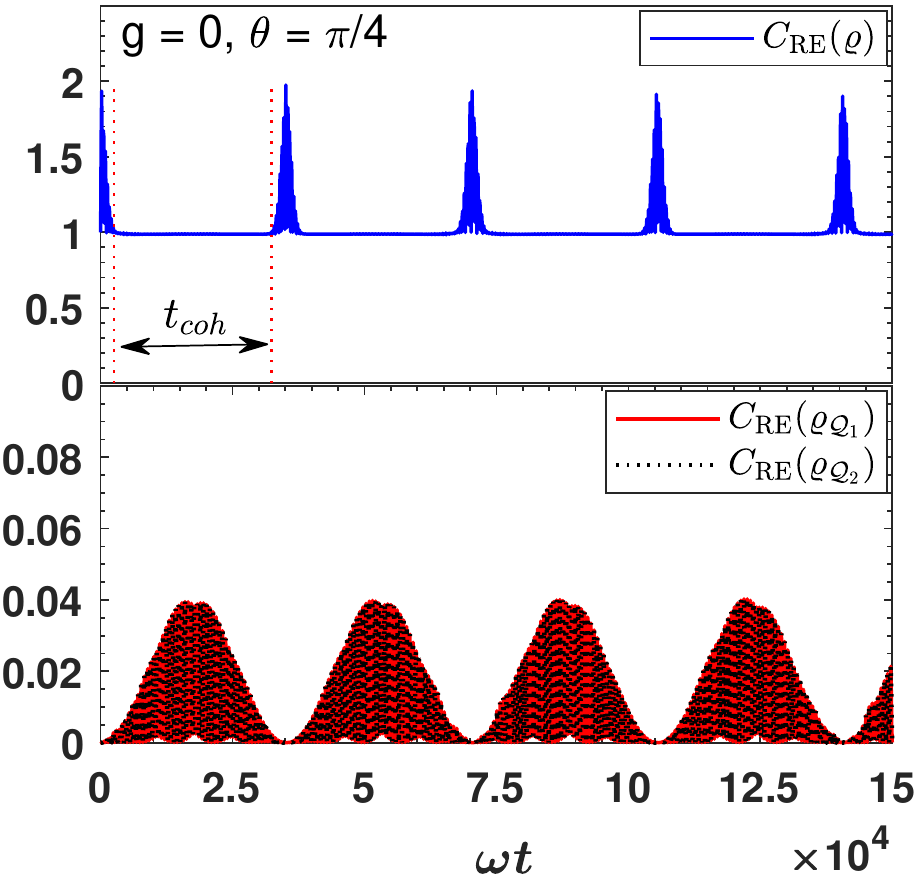}} 
		\captionsetup[subfigure]{labelformat=empty}
		\subfloat[$(\mathsf{c})$]{\includegraphics[width=4.15cm,height=3.8cm]{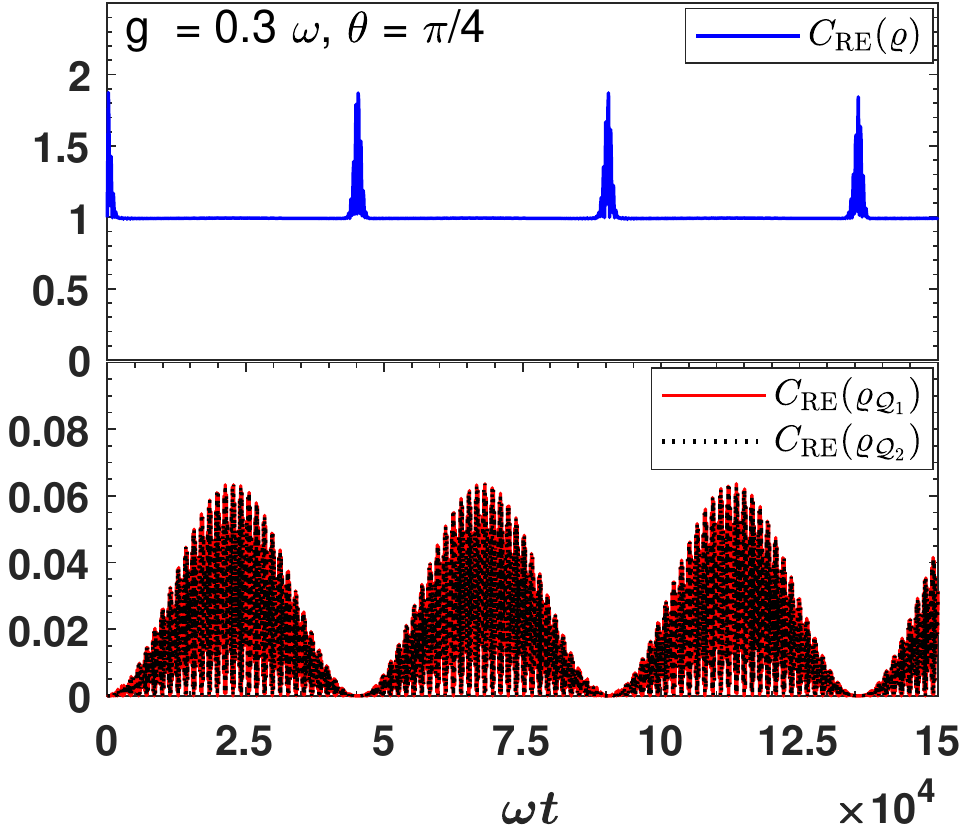}} 
		\captionsetup[subfigure]{labelformat=empty}
		\subfloat[$(\mathsf{d})$]{\includegraphics[width=4.15cm,height=3.8cm]{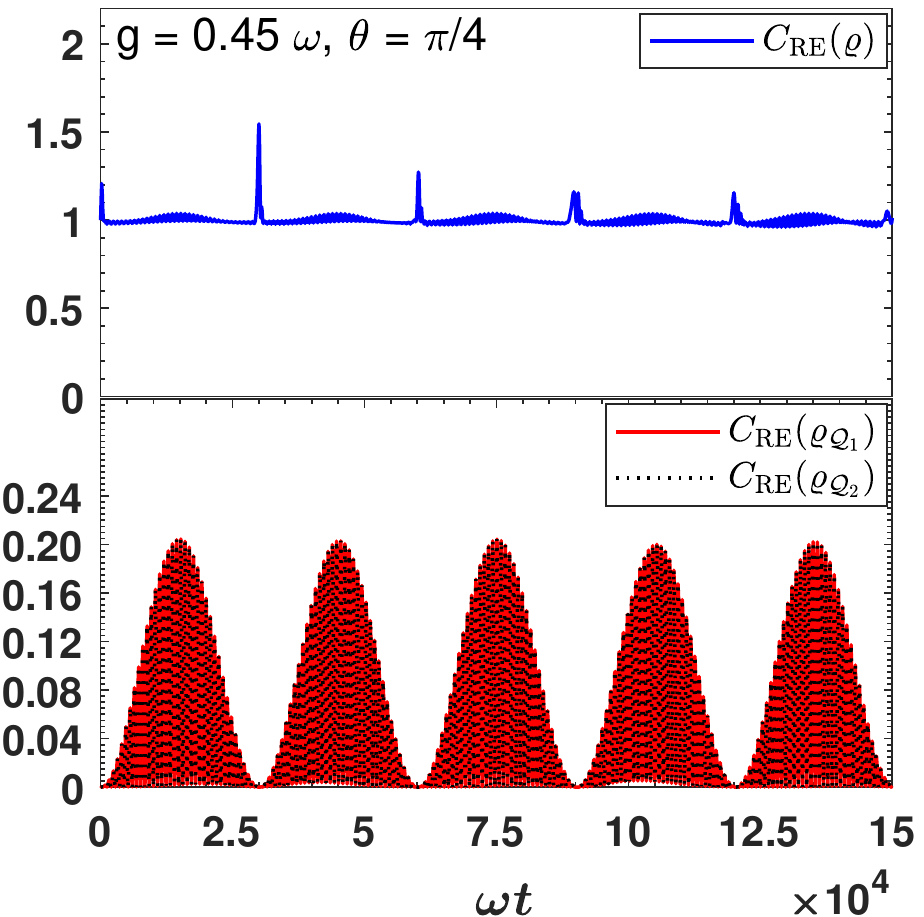}} 
		\caption{ The time evolution of the coherence (\ref{ent_coh}) are shown for $\alpha = 4 $. The blue line indicates the coherence  of the two-qubit subsystem  $(\varrho)$ while the red and black lines illustrate the coherence for  the individual qubit subsystems  $\varrho_{_{\mathcal{Q}_{1}}}$ and $ \varrho_{_{\mathcal{Q}_{2}}}$,  respectively.  Analogous to the case of Fig. \ref{Revival}, here also we consider that both the  qubit's frequencies are same $(0.1 \; \omega)$ and the coupling strength between the qubits and oscillator are equal $(0.015 \; \omega)$. The plot $(\mathsf{a})$ describes the coherence for initially factorized state $(\theta = 0, \; \phi = 0)$ in the absence of parametric term $(g = 0)$.  The plot $(\mathsf{b})$ demonstrates the coherence for initially entangled state $(\theta = \pi/4, \; \phi = 0)$  with $g = 0$. Here, $t_{\text{coh}}$ represents the time interval between the two peaks in $C_{\text{RE}}(\varrho)$. In the presence of parametric oscillator $(g \neq 0)$, the coherence are depicted in $(\mathsf{c})$ and $(\mathsf{d})$.}  
		\label{coh}
	\end{center}
\end{figure}

\section{Relative entropy of coherence}
\label{cohen}

To study the dynamics of quantum coherence in our system, we use  the relative entropy of coherence [\cite{baumgratz2014}], which is defined as 
\beq
C_{\text{RE}}(\rho) = S(\rho_{_{\text{diag}}}) -  S(\rho),
\label{ent_coh}
\eeq
where $ \rho_{_{\text{diag}}} $ is obtained by removing all the off-diagonal elements and keeping the diagonal elements in the density matrix $\rho $. One of the important properties of the relative entropy of coherence is:  $ C_{\text{RE}} (\rho) \leq S(\rho_{_{\text{diag}}}) \leq \log_{2} (d) $,  [\cite{baumgratz2014}],  where $d$ is the dimension of the Hilbert space. It is obvious that if  $ C_{\text{RE}} (\rho) = S(\rho_{_{\text{diag}}})$, the  corresponding quantum state $\rho $ is a pure state. In particular, if there are pure states with $ C_{\text{RE}} (\rho) =  \log_{2} (d) $, these pure states are referred to as maximally coherent states. Note that the term `coherent state' used in this context is not to be confused with the oscillator's coherent  state $\ket{\alpha}$. We use the two-qubit states $\{ \ket {1,1}, \ket{-1,1},\ket{1,-1},\ket{-1,-1}\}$ as the reference basis in our calculation as $C_{\text{RE}}(\rho)$ generally depends on the choice of basis. 

\par
We  consider a simple case which includes the qubit's frequencies are same and the  coupling strength between the qubits and the field are equal. In this case, it is noticed from Fig. \ref{coh}, that the individual qubit subsystems $\varrho_{_{\mathcal{Q}_{1}}}$  and  $\varrho_{_{\mathcal{Q}_{2}}}$ show the equal coherence.  It is evident that the coherence of $\varrho$ is different from that of  $\varrho_{_{\mathcal{Q}_{1}}}$  and  $\varrho_{_{\mathcal{Q}_{2}}}$. For example, in the Fig.  \ref{coh} $(\mathsf{a})$ (in the absence of parametric oscillator, $g = 0$), for the initially factorized state, it is observed that the coherence of $\varrho_{_{\mathcal{Q}_{1}}}$  and  $\varrho_{_{\mathcal{Q}_{2}}}$ encounter null value between the two major revivals whereas the coherence of $\varrho$  sustains its nonzero value. However, this behaviour of coherence changes if we consider the initial state as a maximally  entangled state. For instance, the Fig. \ref{coh} $(\mathsf{b})$ demonstrates  that when $ g = 0$, the coherence of $\varrho$ exhibits nonzero steady value during the time interval $t_{\text{coh}}$. Unlike the case of initially factorized state, the coherence of $\varrho_{_{\mathcal{Q}_{1}}}$  and  $\varrho_{_{\mathcal{Q}_{2}}}$, in this case show the nonvanishing value in the time interval $t_{\text{coh}}$. 
\par
Moreover, it is apparent from the Fig. \ref{coh} $(\mathsf{c})$, that as we increase the parametric strength $g$, the coherence of $\varrho_{_{\mathcal{Q}_{1}}}$  and  $\varrho_{_{\mathcal{Q}_{2}}}$ increase whereas the amplitude of revival peaks in the coherence of $\varrho$ are not greatly affected.  In addition, it is observed that the time interval $t_{\text{coh}}$ also increases with $g$ up to  $g = 0.3 \; \omega $ (Fig. \ref{coh} $(\mathsf{c})$).  However, if we  further increase the parametric strength, say, $0.3\;\omega < g \lesssim 0.45 \; \omega$, the amplitude of revival peaks in the coherence of $\varrho$ gradually decreases as well as  the time interval  $t_{\text{coh}}$ reduces  (Fig. \ref{coh} $(\mathsf{d}))$. We can also observe  that in the limit of $g \to 0.5 \; \omega$, the coherence of  $\varrho $ shows multiple fluctuations and  starts to decrease below the value $1$.
 
\section{Geometric discord and concurrence}
\label{Geo_dis}

To explore the nonclassical correlations that go beyond the entanglement, we utilize the geometric measure of quantum discord [\cite{vedral2010}] defined as
\beq
D_{G}(\rho) = \min_{\wp \in \Omega_{0}} \| \rho - \wp \| ^{2},
\label{dis}
\eeq
where $\Omega_{0}$ denotes the set of zero-discord states. Due to calculational complexity of the  quantum discord [\cite{ollivier2001}], we choose its geometrized version for our two-qubit subsystem. Now, the $4 \times 4 $  density matrix $\varrho$ in the so-called  Bloch basis [\cite{ver2001}] reads as
\bea
\varrho = \frac{1}{4} \left( I \otimes I + \sum_{\imath=1}^{3} 
(a^{\imath} \sigma^{\imath} \otimes I + b^{\imath} I \otimes  \sigma^{\imath} ) 
+ \sum_{\imath,\jmath=1}^{3} \mathcal{T}_{\imath \jmath} \, \sigma^{\imath} \otimes \sigma^{\jmath} \right),
\eea
where  $a^{\imath} = \mathrm{Tr}(\varrho(\sigma^{\imath} \otimes I))$,  and  
$b^{\imath} = \mathrm{Tr}(\varrho( I \otimes \sigma^{\imath} ))$ are  components of the local Bloch vectors,  $\mathcal{T}_{\imath \jmath} = \mathrm{Tr}(\varrho( \sigma^{\imath} \otimes \sigma^{\jmath}))$ are components of the correlation tensor. Therefore,  from (\ref{dis}), it is shown that the geometric measure of quantum discord [\cite{vedral2010}] can be expressed as 
\beq
D_{G}(\varrho)= \frac{1}{4} \left( \norm{a}^{2} + \norm{\mathcal{T}}^{2}- E_{max}\right),
\label{redu_dis}
\eeq
where the column vector $a=(a^{1},a^{2},a^{3})^{\mathrm{T}}$, $\norm{a}^{2}=\sum_{\imath=1}^{3} (a^{\imath})^{2}$, 
$\norm{\mathcal{T}}^{2}= \mathrm{Tr}(\mathcal{T}^{T}\mathcal{T})$ and $E_{max}$ is the largest eigenvalue of the 
matrix $aa^{T}+\mathcal{T}\mathcal{T}^{T}$. Here, the superscript $T$ indicates transpose. In our case, the $D_{G}(\varrho)$ can be evaluated with the following quantities,
\bea
a^{1} &=& 2 \; \mathrm{Re} (\varrho_{_{1,1;1,-1}}+\varrho_{_{-1,1;-1,-1}}), \quad
a^{2}=-2 \; \mathrm{Im} (\varrho_{_{1,1;1,-1}}+\varrho_{_{-1,1;-1,-1}}), \nn \\
a^{3} &=& \varrho_{_{1,1;1,1}} + \varrho_{_{-1,1;-1,1}} - \varrho_{_{1,-1;1,-1}} -\varrho_{_{-1,-1;-1,-1}}, \nn \\
b^{1} &=& 2 \; \mathrm{Re} (\varrho_{_{1,1;-1,1}}+\varrho_{_{1,-1;-1,-1}}), \quad
b^{2}=-2 \; \mathrm{Im} (\varrho_{_{1,1;-1,1}}+\varrho_{_{1,-1;-1,-1}}), \nn \\
b^{3} &=& \varrho_{_{1,1;1,1}} - \varrho_{_{-1,1;-1,1}} + \varrho_{_{1,-1;1,-1}} -\varrho_{_{-1,-1;-1,-1}},
\eea
\beq
\mathcal{T}= 
\begin{pmatrix}
		\mathcal{T}_{11} & 	-2 \mathrm{Im}(\varrho_{_{1,1;-1,-1}}+\varrho_{_{-1,1;1,-1}})  & 2 \mathrm{Re}(\varrho_{_{1,1;1,-1}}-\varrho_{_{-1,1;-1,-1}})  \\
		-2 \mathrm{Im}(\varrho_{_{1,1;-1,-1}}+\varrho_{_{-1,1;1,-1}})& \mathcal{T}_{22} & -2 \mathrm{Im}(\varrho_{_{1,1;1,-1}}-\varrho_{_{-1,1;-1,-1}}) \\
	    2 \mathrm{Re}(\varrho_{_{1,1;-1,1}}-\varrho_{_{1,-1;-1,-1}})
		& -2 \mathrm{Im}(\varrho_{_{1,1;-1,1}}-\varrho_{_{1,-1;-1,-1}}) & \mathcal{T}_{33}		
\end{pmatrix},
\eeq
where the diagonal elements read as:  $\mathcal{T}_{11} = 2 \mathrm{Re}(\varrho_{_{1,1;-1,-1}}+\varrho_{_{-1,1;1,-1}})$,\; $\mathcal{T}_{22}  =  -2 \mathrm{Re}(\varrho_{_{1,1;-1,-1}}-\varrho_{_{-1,1;1,-1}})$, \;  and $\mathcal{T}_{33} =  \varrho_{_{1,1;1,1}} - \varrho_{_{-1,1;-1,1}} - \varrho_{_{1,-1;1,-1}} + \varrho_{_{-1,-1;-1,-1}}$.  It is clear from (\ref{redu_dis}) that $D_{G}(\varrho)$ is not normalized to unity, its maximum  value is 1/2. Hence, we consider $2D_{G}(\varrho)$ to be a proper measure [\cite{girolami2011}] for comparison with the concurrence.

\begin{figure}
	\begin{center}
		\captionsetup[subfigure]{labelformat=empty}
		\subfloat[$(\mathsf{a})$]{\includegraphics[width=5cm,height=4cm]{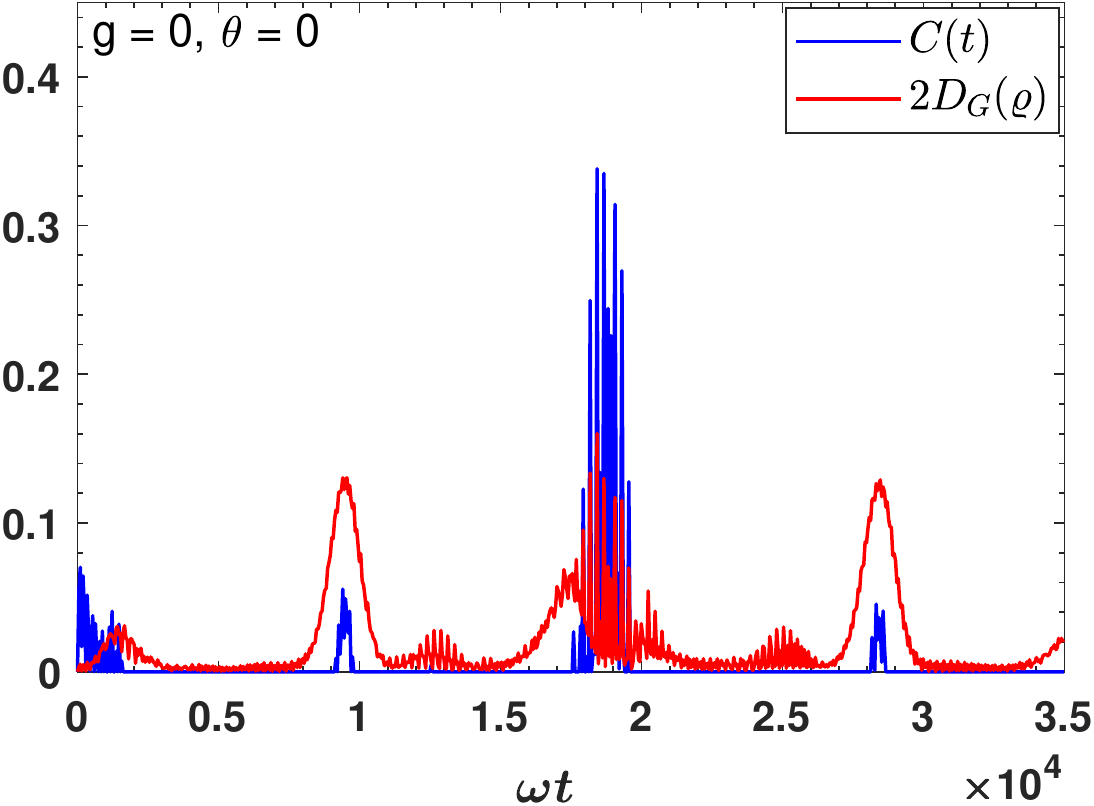}} 
		\captionsetup[subfigure]{labelformat=empty}
		\subfloat[$(\mathsf{b})$]{\includegraphics[width=5cm,height=4cm]{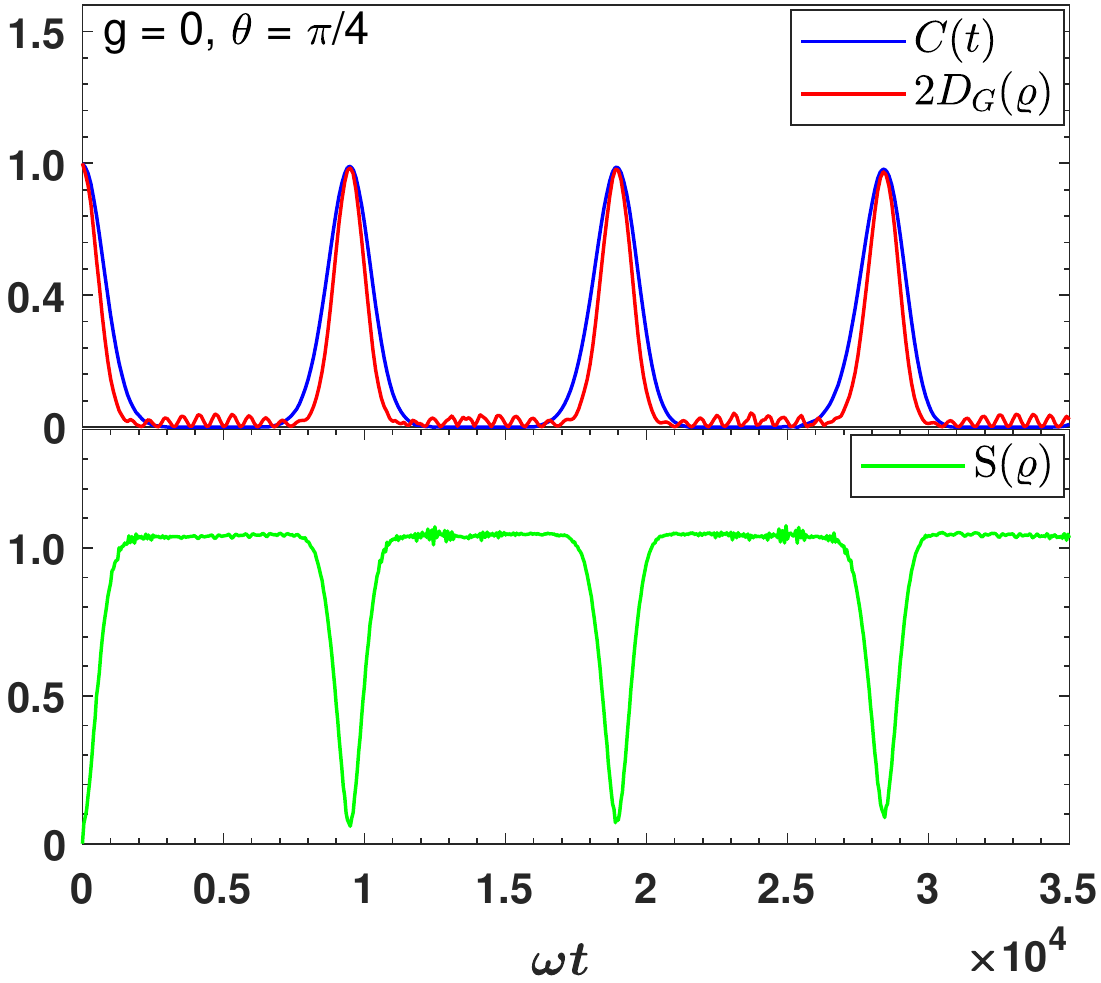}} 
		\captionsetup[subfigure]{labelformat=empty}
		\subfloat[$(\mathsf{c})$]{\includegraphics[width=5cm,height=4cm]{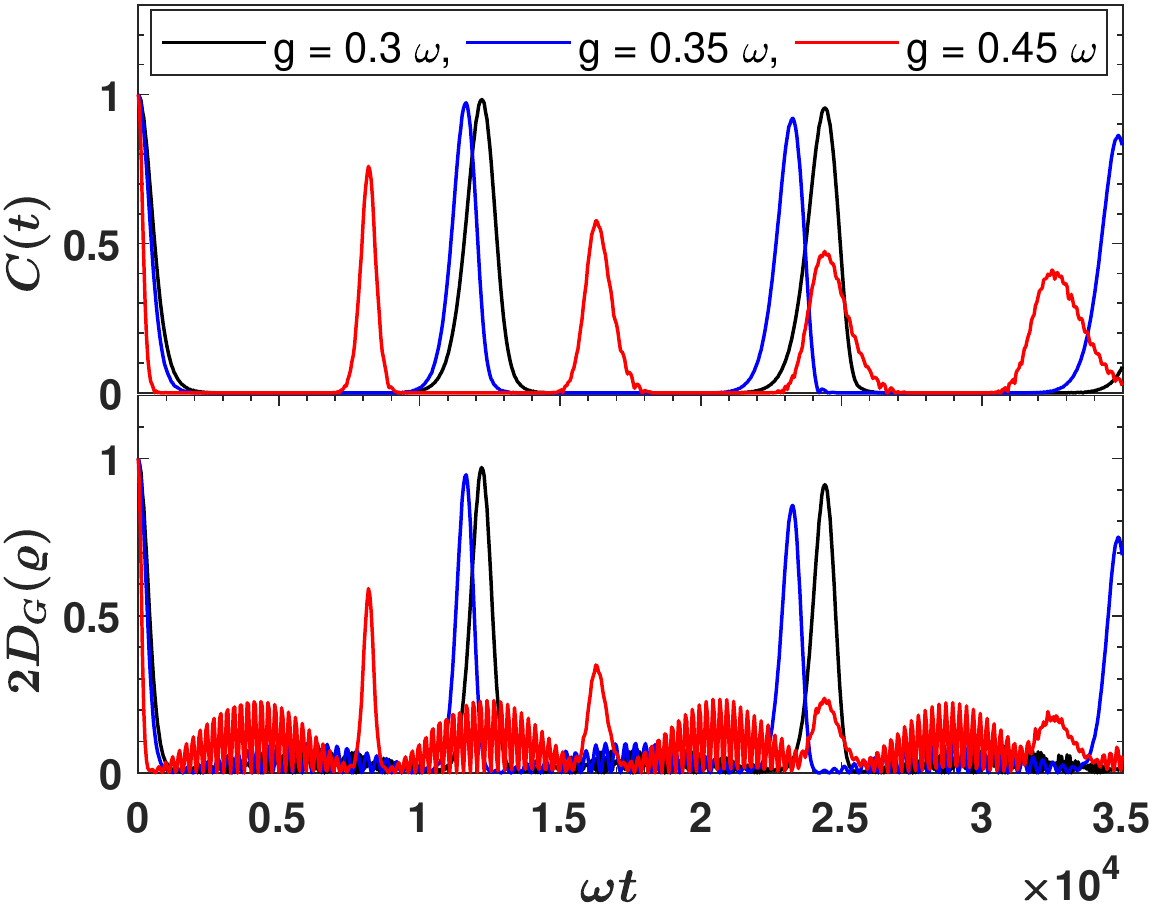}} 
		\captionsetup[subfigure]{labelformat=empty}
		\caption{  The time evolution of $2D_{G}(\varrho)$, $C(t)$ and Von Neumann entropy $(\mathrm{S}(\varrho))$ are presented for the qubit's frequencies $\Delta_{1}=0.1\; \omega$, and $ \Delta_{2}=0.08 \; \omega$ with $\alpha =2$.  Here, the coupling strength $(\lambda_{1},\; \lambda_{2})$ are considered to be $(0.02\; \omega, \; 0.04 \; \omega)$  for different values of the parametric strength  $g$. The plot $(\mathsf{a})$ describes the behaviour  of $ 2D_{G}(\varrho)$ and $C(t)$ in the absence of parametric term $(g = 0)$ for initially factorized state $(\theta = 0, \; \phi = 0)$. The plots $(\mathsf{b})$ and $(\mathsf{c})$ illustrate the same for initially entangled state $(\theta = \pi/4,\; \phi = 0)$. The green line in $(\mathsf{b})$ depicts the evolution of $\mathrm{S}(\varrho)$. }  
		\label{con_dis}
	\end{center}
\end{figure}

\par
We use the concurrence which is  widely accepted to determine the degree of entanglement between the qubits. The concurrence [\cite{wootters1998}] is defined as
\beq
C(t) = \max \big\{0,\sqrt{\lambda_{1}}-\sqrt{\lambda_{2}}-\sqrt{\lambda_{3}}-\sqrt{\lambda_{4}}\big\},
\label{concurrence}
\eeq
where $\big\{\lambda_{\imath}|\, \imath = (1, \ldots, 4)\big\}$ are the eigenvalues arranged in descending order of the matrix 
\beq
{\mathsf R}(t) = \varrho(t) \widetilde{\varrho}(t), \qquad
\widetilde{\varrho}(t) = \left( \sigma^{\mathrm {y}}  \otimes \sigma^{\mathrm{y}} \right)  \varrho(t)^{*}  \left( \sigma^{\mathrm {y}}  \otimes \sigma^{\mathrm{y}} \right).
\label{R_t}
\eeq
The matrix  $\widetilde{\varrho}(t)$ is obtained under the spin-flip operation on the  two-qubit reduced density matrix  $\varrho(t)$. The two-qubit subsystem shows entanglement  for $C(t) > 0$. The maximum value of entanglement can be achieved when $C(t) = 1$, while $ C(t) = 0$ implies separability.
\par
It is already reported that the absence of entanglement between a pair of systems does not imply classicality i.e.  there may be nonvanishing quantum correlations which are  measured by quantum discord [\cite{ollivier2001}]. In our case, we  demonstrate the behaviour of  $ 2D_{G}(\varrho)$ and $C(t)$ and their dependence on the parametric strength $g$. It is observed that  $ 2D_{G}(\varrho)$ shows small amplitude oscillations in the  entanglement sudden death region both for initially factorized $(\theta = 0,\; \phi = 0)$ as well as maximally entangled states $(\theta = \pi/4,\; \phi = 0)$ (Fig. \ref{con_dis}). Thus it is evident even if the entanglement between the two qubits vanishes, there are still quantum correlations between them.   It is noticed that for initial state as a  maximally entangled state with $ g =0 $ (Fig. \ref{con_dis} $(\mathsf{b})$), the evolved state becomes mixed state in the entanglement sudden death region and tending towards pure state when $C(t)$ reaches its maximum value. As we increase $g$, the time interval of entanglement sudden death decreases whereas the amplitude of oscillations in $ 2 D_{G}(\varrho)$ increases which are obvious from the Fig. \ref{con_dis} $(\mathsf{c})$. Moreover, we calculate $ C(t)$ for the reconstruction of  the evolved state for $\varrho$ in the minimum entropy configuration which is discussed in Sec. \ref{nonclassical}. 

\section{Generation of nonclassical states}
\label{nonclassical}
\subsection{The generalized Bell states }

\begin{figure}
	\begin{center}
		\captionsetup[subfigure]{labelformat=empty}
		\subfloat[$(\mathsf{a})$]{\includegraphics[width=6cm,height=5cm]{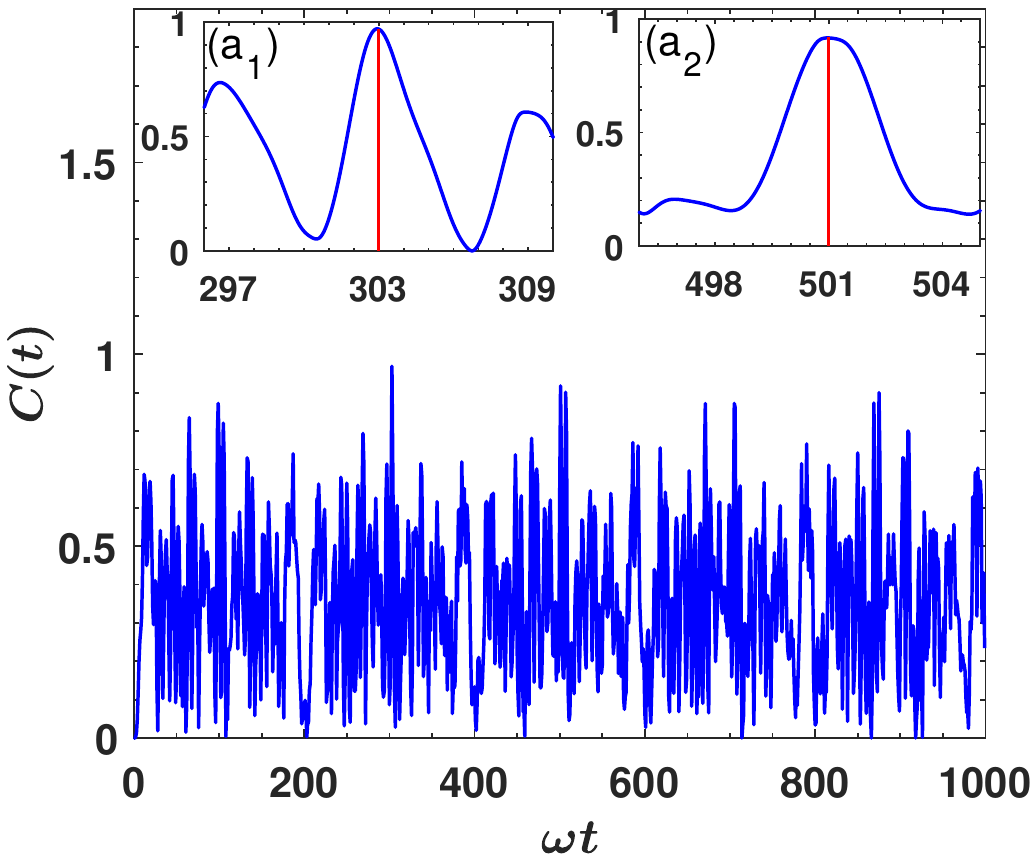}} 
		\captionsetup[subfigure]{labelformat=empty}
		\caption{  The plot $(\mathsf{a})$ describes the time evolution of $C(t)$ for the values of the qubit's frequencies $\Delta_{1}=0.2\; \omega$, and $ \Delta_{2}=0.15 \; \omega$ with $\alpha  = 0$. The coupling strength $(\lambda_{1}, \lambda_{2})$ in this case are chosen to be $(0.32 \; \omega,\; 0.17\; \omega)$. The initial state is chosen as a factorized state $(\theta = 0,\; \phi =0)$ with $g = 0$. Two vertical red lines  drawn at  $\omega t = 303 $ and  $ 501 $ in the inset $(\mathsf{a}_{1})$ and $(\mathsf{a}_{2})$ respectively, indicate the nearly maximal value of  $C(t)$.}  
		\label{recon}
	\end{center}
\end{figure}

We now examine the evolution of the entanglement of the two-qubit reduced density matrix  $ \varrho $  (\ref{rdens}) as quantified by $ C(t) $ in (\ref{concurrence}). It is observed that the dynamical evolution produces the superposition of two-qubit states which are in close proximity to the maximally entangled generalized Bell states. For this, we consider the  $ \varrho $ at specific times where the  concurrence achieves its  maximum value  i.e. $ C(t)\lesssim 1 $. For instance, at the times $\omega t = 303$ and $501$, the values of $C(t)$ read as $0.96$ and $0.92$ respectively which are shown in the inset of  Fig. \ref{recon} ($(\mathsf{a}_{1})$  and  $(\mathsf{a}_{2})$). To determine the states at those aforesaid times, we compute the Hilbert-Schmidt distance $(\mathrm{d}_{\mbox{\tiny{HS}}})$ [\cite{dodonov2000}] between $\varrho$ and a pure state density matrix $\rho'$: 

\beq
\mathrm{d}_{\mbox{\tiny{HS}}} = \sqrt{\mathrm{Tr}\,(\varrho - \varrho^{'})^{2}}, \;
\varrho{'} = \ket{\Upsilon}\bra{\Upsilon}, \; 
\ket{\Upsilon} = \alpha_{1} \ket{\Phi_{+}}+\alpha_{2} \ket{\Phi_{-}} +\alpha_{3} \ket{\Psi_{+}} + \alpha_{4} \ket{\Psi_{-}},
\label{hilbert}
\eeq
where the generalized Bell basis states read as
\beq
\ket{\Phi_{\pm}} = \dfrac{1}{\sqrt{2}} (\ket{1,1} \pm i \ket{-1,-1}), \quad
\ket{\Psi_{\pm}} = \dfrac{1}{\sqrt{2}} (\ket{1,-1} \pm i \ket{-1,1}).
\label{gen_bell}
\eeq
\begin{table}[H]
	\centering
	\begin{tabular}{|c|M{1.7in}|M{2.35in}|}
		\cline{1-3}
		\multicolumn{1}{|M{0.5in}|}{$\omega t$} &    $303$         &     $501$               \\ \hline
		\multicolumn{1}{|M{0.5in}|}{$C(t)$}    &      $0.96$       &       $0.92$             \\ \hline
		\multicolumn{1}{|M{0.5in}|}{$\mathrm{Tr}(\varrho(t)^{2})$}    &      $0.97$       &       $0.93$   \\ \hline
		\multicolumn{1}{|M{0.5in}|}{$\mathrm{d}_{\mbox{\tiny{HS}}} \lvert_{\min} $}    &   $0.04$          &    $0.07$            \\ \hline
		\multicolumn{1}{|M{0.5in}|}{$ \ket{\Upsilon} $}  & \parbox{1.7in}{\centering{$0.995407 \ket{\Phi_{-}} - $  $(0.039816 $ \\ $- 0.054747 i)$ $ (\ket{\Psi_{+}} - \ket{\Psi_{-}} )$  \\     $ \approx \ket{\Phi_{-}}$}}  &    \parbox{2.3in}{\centering{ $0.969071 \ket{\Phi_{+}}-$ $(0.02900 - 0.18896 i)$  $ \ket{\Psi_{+}} -(0.125979 $ $+0.0920618 i)\ket{\Psi_{-}}$   \\ $ \approx \ket{\Phi_{+}}$}}         \\ \hline
	\end{tabular}
	\caption{}
	\label{tab}
\end{table}

Note that the  above coefficients $(\alpha_{\imath} \in \mathbb{C})$  satisfy  the normalization condition i.e. $\sum_{\imath=1}^{4}|\alpha_{\imath}|^{2} =1$. In the numerical minimization  procedure, these coefficients are altered  to find a suitable linear combination  of the generalized Bell states (\ref{gen_bell}) that minimizes the $\mathrm{d}_{\mbox{\tiny{HS}}}$ (\ref{hilbert}) over the ensemble of states $\{\ket{\Upsilon} \mid (\alpha_{1}, \alpha_{2}, \alpha_{3}, \alpha_{4}) \; \in \mathbb{C}\}$. In this case, we adopt the initial state as a factorized state $(\theta = 0, \; \phi = 0)$ to emphasize the dynamical effects that give rise to  nearly pure entangled states at the times mentioned in the inset of the  Fig. \ref{recon} $(\mathsf{a})$.  The relevant quantities and the characterization  of states that show minimum  $\mathrm{d}_{\mbox{\tiny{HS}}}$ are inscribed in the Table \ref{tab}. For our chosen parameters, it is observed that at  $ \omega t = 303$, where $ C(t)$ reaches its nearly maximal value, the resultant two-qubit density matrix predominantly behaves as one of the generalized Bell state density matrix $\varrho \sim \ket{\Phi_{-}} \bra {\Phi_{-}}$ with purity $97\%$. Similarly, at later time $\omega t= 501 $, it is found that the resultant $\varrho$ is greatly governed by another generalized Bell state density matrix i.e. $\varrho \sim \ket{\Phi_{+}} \bra {\Phi_{+}}$ having purity $93\%$ which is shown explicitly in the Table \ref{tab}.

\begin{figure}
	\begin{center}
		\captionsetup[subfigure]{labelformat=empty}
		\subfloat[$(\mathsf{a})$]{\includegraphics[width=5.5cm,height=5.cm]{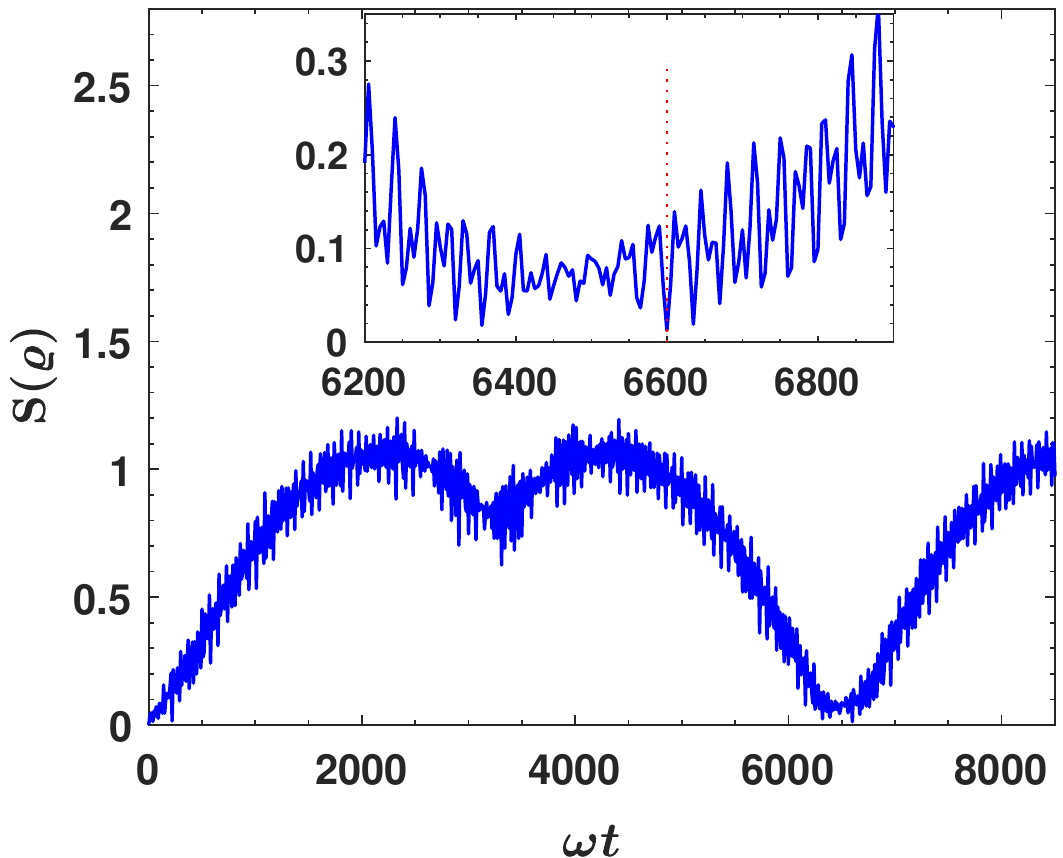}} 
		\captionsetup[subfigure]{labelformat=empty}
		\captionsetup[subfigure]{labelformat=empty}
		\subfloat[$(\mathsf{b})$]{\includegraphics[width=6.5cm,height=5.cm]{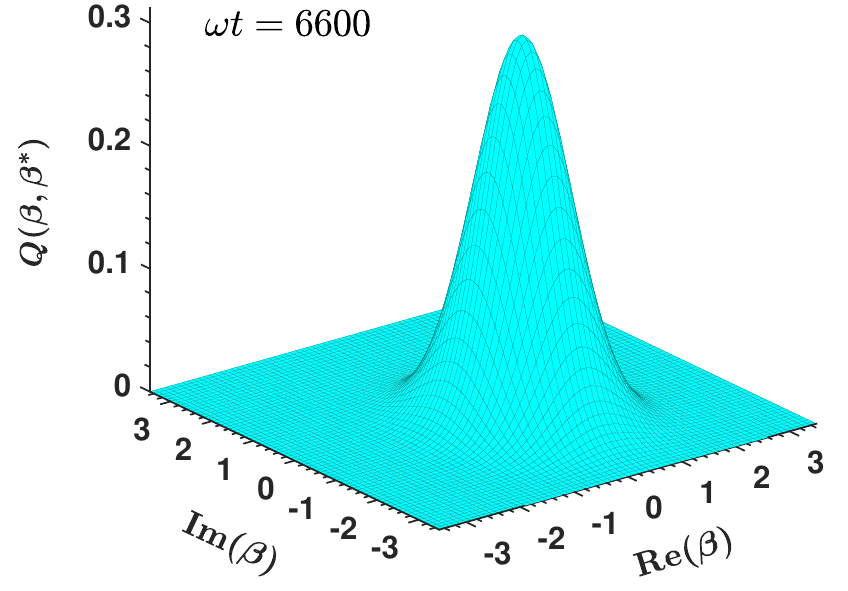}} 
		\captionsetup[subfigure]{labelformat=empty}
		\caption{$(\mathsf{a})$ The time evolution of  $\mathrm{S}(\varrho)$ is presented in the presence of parametric strength $(g = 0.1\; \omega )$ for the equal qubit's frequencies $(\Delta_{1}, \Delta_{2})$ and equal coupling strength $(\lambda_{1}, \lambda_{2})$ with the values $ 0.08\; \omega $ and $ 0.06\; \omega $, respectively. Here, the initial field state is taken as a coherent state $\ket{\alpha}$  with $\alpha = 0.5$ whereas the two-qubit state is a factorized state $(\theta = 0,\; \phi = 0)$. In the inset, the dotted-vertical line is drawn at $\omega t = 6600$ where the value of $\mathrm{S}(\varrho)$  is obtained as  $0.0125$. The plot $(\mathsf{b})$ shows the Husimi $Q$-function on the phase-space at time $\omega t = 6600$ for the parameter values as of \textbf{$(\mathsf{a})$}. }  
    \label{SCS}
	\end{center}
\end{figure} 
\subsection{The squeezed coherent states}
Finally, we investigate the generation of squeezed coherent state corresponding  to the oscillator degree of freedom at the minimum entropy regime. We adopt the initial state of the field as the coherent state, $\ket{\alpha}$ with $\alpha = 0.5$ while the two-qubit state is considered as a factorized state $(\theta = 0, \; \phi = 0)$.  We study the quadrature squeezing and the phase-space distribution of the Husimi Q-function [\cite{schleich2011}]  which enable us to identify the evolved state of the corresponding density matrix $\rho_{\mathcal{O}}(t)$. The principal-quadrature squeezing [\cite{lukvs1988,luks1988acta,mir2010,ma2011}] is characterized by 
\beq
V_{\text{min}}= \min_{\phi \in (0,2\pi)}\limits (\braket{X^{2}_{\phi}}-\braket{X_{\phi}}^{2}) = \frac{1}{2} + \; \braket{a^{\dagger} a} - \; |\braket{a}|^{2} - \; \lvert\braket{a^{2}} -\braket{a}^{2}\rvert,
\label{PVariance}
\eeq
where the quadrature operator is defined as $X_{\phi} = \frac{1}{\sqrt{2}} (a \exp{(-i \phi)} + a^{\dagger} \exp{(i \phi)})$, and $\phi$ is a real phase [\cite{barnett2002}]. The variance $V_{\text{min}}$ (\ref{PVariance}) is equal to 0.5 for both the vacuum and coherent states, which is known as the classical limit of the variance. The state of the field is said to be squeezed [\cite{mandel1982}] if the corresponding variance is less than 0.5. The expectation values of the  operators involved in (\ref{PVariance}) are  explicitly  shown  in the Appendix \ref{appendix:Ap2}.
\par 
The Husimi Q-function is a quasi probability distribution that is defined as the expectation value of the oscillator density matrix in an arbitrary coherent state $\ket{\beta}$. In comparison to the other phase-space quasi probability distributions, it assumes nonnegative values on the phase-space. It has been widely used in the study of occupation on the phase space owing to its ease of computation [\cite{sugita2002,Ingold2003}]. For our reduced density matrix of the oscillator $\rho_{\mathcal{O}}(t)$, the corresponding $Q$-function reads as
\bea
Q(\beta,\beta^{*})  \!\!\! &= & \!\!\! \frac{1}{\pi} \bra\beta\rho_{\cal O}\ket\beta \nn \\
 \!\!\!\! &=& \!\!\!\! \frac{1}{\pi} \sum_{n,m=0}^{\infty} \left( \mathcal{F}_{n,m}^{(1)} \braket{\beta | r,n_{1,1}} \braket{r,m_{1,1}| \beta} + \mathcal{F}_{n,m}^{(-1)} \braket{\beta | r,n_{-1,-1}} \braket{r,m_{-1,-1}|\beta}
\right. \quad \quad \nn \\
& & + \; \mathcal{F}_{n,m}^{(2)} \braket{\beta | r,n_{-1,1}} \braket{r,m_{-1,1}|\beta} +\mathcal{F}_{n,m}^{(-2)} \braket{\beta | r,n_{1,-1}} \braket{r,m_{1,-1} | \beta}\Big).
\label{Q_defn}
\eea
It can be shown that the expression (\ref{Q_defn}) meets the normalization condition i.e. $\int Q(\beta, \beta^{*}) \mathrm{d}^{2}\beta = 1$ and also maintains the bounds: $0 \leq Q(\beta,\beta^{*}) \leq \frac{1}{\pi}$.  The inner products in the above Eq. (\ref{Q_defn}) are calculated using the expression (\ref{inner_product}).
\par
It is already mentioned that $\mathrm{S}(\rho_{\mathcal{O}}) = \mathrm{S}(\varrho)$  as our total system resides in a pure state [\cite{araki2002}]. From the time evolution of $\mathrm{S}(\varrho)$  (Fig. \ref{SCS} $(\mathsf{a})$), it is seen  that  the $\mathrm{S}(\varrho)$ reaches the minimum value  $0.0125$ at the time $\omega t = 6600$ for our chosen values of the parameters. We observe that the value of $V_{\text{min}}$ (\ref{PVariance}) at $\omega t =6600$ reads as $0.3411$ which implies that the evolved state at that time becomes squeezed, as well as  it is noticed that the single peak of the  $Q$-function (\ref{Q_defn}) is displaced from the origin in the $\beta$ phase-space (Fig. \ref{SCS} $(\mathsf{b})$). This indicates that the obtained nearly pure evolved state is a squeezed coherent state.
\section{Conclusion}
\label{sec_con}

Using the adiabatic approximation, we have studied the quantum properties in a strongly interacting system consisting of two-qubit and oscillator in the presence of a parametric oscillator. To validate our approximation, a comparison of the analytically obtained approximate energy spectrum with the numerically calculated spectrum of the entire Hamiltonian is shown. It is observed that when the parametric oscillator's strength reaches a critical value, the excited energy levels of the Hamiltonian start to collapse together which is explained in terms of phase-space variables. From the time evolution of the initially tripartite state, the reduced density matrices of the qubits and the oscillator degrees of freedom are computed.  The effect of  parametric oscillator is demonstrated on  the revival and collapse phenomenon exhibiting in the two-qubit population inversion. 
\par
Considering the initial state as factorized as well as maximally entangled states, we have investigated the quantum coherence for the two-qubit subsystem and its individual qubit subsystems. It is  noticed that the time interval between the coherence peaks for the two-qubit subsystem increases with the parametric oscillator's strength up to a certain value. Similarly,  the behaviour of geometric discord and concurrence  are examined with respect to the two-qubit evolved state, and it is found that there is nonzero geometric discord in the entanglement sudden death region. Moreover, by adopting the initial state as a factorized state, we have shown the creation of generalized Bell states via minimizing the corresponding Hilbert-Schmidt distance. Besides, computing the quadrature variance for the oscillator degree of freedom and observing the phase-space distribution of the corresponding $Q$-function at the  minimum entropy regime, it is concluded that  the nearly pure evolved state is a squeezed coherent state.

\section*{Acknowledgement}
We would like to thank M. Sanjay Kumar for his encouragement and support. One of us (PM) acknowledges the financial support from DST (India) through the INSPIRE Fellowship Programme.
 
\begin{appendices}
\section{}
\label{appendix:Ap1}
\addtocounter{section}{0}
\setcounter{equation}{0}
\numberwithin{equation}{section}
The explicit form of $\varpi(t)$  in (\ref{rho_qubit-osc}) reads as
\bea
\varpi(t) \!\!\!\! &=& \!\!\!\! \sum_{n,m=0}^{\infty} \left( \mathcal{F}_{n,m}^{(1)} \ket{1;r,n_{1,1}} \bra{1;r,m_{1,1}}
+ \mathcal{F}_{n,m}^{(-1)} \ket{-1;r,n_{-1,-1}} \bra{-1;r,m_{-1,-1}}
\right. \quad \quad \nn \\
&+& \!\!\!\! \mathcal{F}_{n,m}^{(2)} \ket{-1;r,n_{-1,1}} \bra{-1;r,m_{-1,1}} +\mathcal{F}_{n,m}^{(-2)} \ket{1;r,n_{1,-1}} \bra{1;r,m_{1,-1}}+ \mathcal{G}_{n,m}^{(1)} \ket{1;r,n_{1,1}} \bra{-1;r,m_{-1,1}} \quad \quad
\nn \\
&+& \!\!\!\!   \mathcal{G}_{n,m}^{(-1)} \ket{-1;r,n_{-1,-1}} \bra{1;r,m_{1,-1}} 
+ \mathcal{G}_{n,m}^{(2)} \ket{-1;r,n_{-1,1}} \bra{1;r,m_{1,1}} \nn \\
&+&  \!\!\!\!  \left. \mathcal{G}_{n,m}^{(-2)} \ket{1;r,n_{1,-1}} \bra{-1;r,m_{-1,-1}} \right),
\label{rdens_qo}
\eea
where the various coefficients are written as
\bea
\mathcal{F}_{n,m}^{(\pm 1)} \!\!\!\! &=& \!\!\!\! \left( \widetilde{\mathcal{C}}_{1,n}^{+}(t) \pm 
\widetilde{\mathcal{C}}_{2,n}^{+}(t) \right) \left( \widetilde{\mathcal{C}}_{1,m}^{+}(t)^{*} \pm 
\widetilde{\mathcal{C}}_{2,m}^{+}(t)^{*} \right), \nn \\
\mathcal{F}_{n,m}^{(\pm 2)} \!\!\!\! &=& \!\!\!\! 
\Big( \frac{\Gamma_{+,n}}{|\Gamma_{+,n}|} \; \widetilde{\mathcal{C}}_{1,n}^{-}(t) \pm
\frac{\Gamma_{-,n}}{|\Gamma_{-,n}|} \; \widetilde{\mathcal{C}}_{2,n}^{-}(t) \Big) 
\Big ( 
\frac{\Gamma_{+,m}}{|\Gamma_{+,m}|} \, 
\widetilde{\mathcal{C}}_{1,m}^{-}(t)^{*} \pm 
\frac{\Gamma_{-,m}}{|\Gamma_{-,m}|} \, 
\widetilde{\mathcal{C}}_{2,m}^{-}(t)^{*} \Big), \nn \\
\mathcal{G}_{n,m}^{(\pm 1)} \!\!\!\! &=& \!\!\!\! \left( \widetilde{\mathcal{C}}_{1,n}^{+}(t) \pm 
\widetilde{\mathcal{C}}_{2,n}^{+}(t) \right) 
\Big ( 
\frac{\Gamma_{+,m}}{|\Gamma_{+,m}|} \, 
\widetilde{\mathcal{C}}_{1,m}^{-}(t)^{*} \pm 
\frac{\Gamma_{-,m}}{|\Gamma_{-,m}|} \, 
\widetilde{\mathcal{C}}_{2,m}^{-}(t)^{*} \Big), \nn \\
\mathcal{G}_{n,m}^{(\pm 2)} \!\!\!\! &=& \!\!\!\! \left( \widetilde{\mathcal{C}}_{1,m}^{+}(t)^{*} \pm 
\widetilde{\mathcal{C}}_{2,m}^{+}(t)^{*} \right)
\Big( \frac{\Gamma_{+,n}}{|\Gamma_{+,n}|} \; \widetilde{\mathcal{C}}_{1,n}^{-}(t) \pm
\frac{\Gamma_{-,n}}{|\Gamma_{-,n}|} \; \widetilde{\mathcal{C}}_{2,n}^{-}(t) \Big).
\eea
\section{}
\renewcommand{\thesection}{B}
\addtocounter{section}{0}
\setcounter{equation}{0}
\numberwithin{equation}{section}
\label{appendix:Ap2}
The following expectation values are utilized to calculate the quadrature variance (\ref{PVariance}).
\bea
\braket{a^{k}}=\sum_{n,m=0}^{\infty} 
\left( \mathcal{F}_{n,m}^{(1)}(t) \mathcal{H}_{n,m}^{(1,k)}(t) + 
\mathcal{F}_{n,m}^{(-1)}(t) \mathcal{H}_{n,m}^{(-1,k)}(t) 
+ \mathcal{F}_{n,m}^{(2)}(t) \mathcal{H}_{n,m}^{(2,k)}(t) +
\mathcal{F}_{n,m}^{(-2)}(t) \mathcal{H}_{n,m}^{(-2,k)}(t)\right),
\eea
where,
\bea
\mathcal{H}_{n,m}^{(\pm1,k)}(t) \!\!\! &=& \!\!\! \sum_{\ell=k}^{\infty} \sqrt{\tfrac{\ell!}{(\ell-k)!}}
\braket{\ell|\mathrm{S}^{\dagger}(r)\mathrm{D}^{\dagger}(\eta_{\pm1,\pm1})|n}
\braket{m|\mathrm{D}(\eta_{\pm1,\pm1})\mathrm{S}(r)|\ell-k}, \nn \\
\mathcal{H}_{n,m}^{(\pm2,k)}(t) \!\!\! &=& \!\!\! \sum_{\ell=k}^{\infty} \sqrt{\tfrac{\ell!}{(\ell-k)!}}
\braket{\ell|\mathrm{S}^{\dagger}(r)\mathrm{D}^{\dagger}(\eta_{\mp1,\pm1})|n}
\braket{m|\mathrm{D}(\eta_{\mp1,\pm1})\mathrm{S}(r)|\ell-k},
\eea
\bea
\braket{a^{\dagger}a}=\sum_{n,m=0}^{\infty} 
\left( \mathcal{F}_{n,m}^{(1)}(t) \mathcal{J}_{n,m}^{(1)}(t) + 
\mathcal{F}_{n,m}^{(-1)}(t) \mathcal{J}_{n,m}^{(-1)}(t) 
+ \mathcal{F}_{n,m}^{(2)}(t) \mathcal{J}_{n,m}^{(2)}(t) +
\mathcal{F}_{n,m}^{(-2)}(t) \mathcal{J}_{n,m}^{(-2)}(t)\right),
\eea
where,
\bea
\mathcal{J}_{n,m}^{(\pm1)}(t) \!\!\! &=& \!\!\! \sum_{\ell=0}^{\infty} \ell
\braket{\ell|\mathrm{S}^{\dagger}(r)\mathrm{D}^{\dagger}(\eta_{\pm1,\pm1})|n}
\braket{m|\mathrm{D}(\eta_{\pm1,\pm1})\mathrm{S}(r)|\ell}, \nn \\
\mathcal{J}_{n,m}^{(\pm2)}(t) \!\!\! &=& \!\!\! \sum_{\ell=0}^{\infty} \ell
\braket{\ell|\mathrm{S}^{\dagger}(r)\mathrm{D}^{\dagger}(\eta_{\mp1,\pm1})|n}
\braket{m|\mathrm{D}(\eta_{\mp1,\pm1})\mathrm{S}(r)|\ell},
\eea
\bea
\bra{m}D(\alpha) S(\xi)\ket{n} &=& \sqrt{\frac{m!}{n! \mu}} i^{m} \bigg(\frac{\nu}{2\mu}\bigg)^{\frac{m}{2}} \exp{\bigg(-\frac{\lvert \alpha \rvert ^2}{2} +\frac{\alpha^{*2} \nu}{2 \mu}}\bigg) \sum \limits_{k=0}^{\text{min} (m,n)} \binom {n}{k} \frac{(-i)^k}{(m-k)!} \times \nn \\
&\times & \bigg( \frac{2}{\mu \nu}\bigg)^{\frac{k}{2}} \bigg(\frac{\nu^{\ast}}{2 \mu}\bigg)^{\frac{n-k}{2}} \; \mathrm{H}_{\text{n-k}} \bigg(\frac{-\alpha^{*}}{\sqrt{2 \mu \nu^{*}}}\bigg)  \mathrm{H}_{\text{m-k}} \bigg(\frac{-i\big( \mu \alpha - \nu \alpha^{*}\big)}{\sqrt{2 \mu \nu}}\bigg).
\label{inner}
\eea
\end{appendices}


\end{document}